# The Impact of Extended CO$_2$ Cross Sections on Temperate Anoxic Planet Atmospheres

Wynter Broussard[1], Edward W. Schwieterman[1,2], Clara Sousa-Silva[3,4], Grace Sanger-Johnson[5], Sukrit Ranjan[2,6], and Olivia Venot[7]
[1] Department of Earth and Planetary Sciences, University of California, Riverside, CA 92521, USA; abrou009@ucr.edu
[2] Blue Marble Space Institute of Science, Seattle, WA 98104, USA
[3] Bard College, 30 Campus Rd, Annandale-On-Hudson, NY 12504, USA
[4] Institute of Astrophysics and Space Sciences, Rua das Estrelas, 4150-762 Porto, Portugal
[5] Department of Physics and Astronomy, Michigan State University, East Lansing, MI 48824, USA
[6] University of Arizona, Lunar and Planetary Laboratory/Department of Planetary Sciences, Tucson, AZ 85721, USA
[7] Université Paris Cité and Univ. Paris Est Creteil, CNRS, LISA, F-75013 Paris, France
Received 2024 November 20; revised 2025 January 13; accepted 2025 January 14; published 2025 February 14

## Abstract

Our interpretation of terrestrial exoplanet atmospheric spectra will always be limited by the accuracy of the data we use as input in our forward and retrieval models. Ultraviolet molecular absorption cross sections are one category of these essential model inputs; however, they are often poorly characterized at the longest wavelengths relevant to photodissociation. Photolysis reactions dominate the chemical kinetics of temperate terrestrial planet atmospheres. One molecule of particular importance is CO$_2$, which is likely present in all terrestrial planet atmospheres. The photolysis of CO$_2$ can introduce CO and O, as well as shield tropospheric water vapor from undergoing photolysis. This is important because H$_2$O photolysis produces OH, which serves as a major reactive sink to many atmospheric trace gases. Here, we construct CO$_2$ cross-section prescriptions at 195 K and 300 K extrapolated beyond 200 nm from measured cross sections. We compare results from the implementation of these new cross sections to the most commonly used CO$_2$ prescriptions for temperate terrestrial planets with Archean-like atmospheres. We generally find that the observational consequences of CO$_2$ dissociation beyond 200 nm are minimal so long as our least conservative (highest opacity) prescription can be ruled out. Moreover, implementing our recommended extended CO$_2$ cross sections does not substantially alter previous results that show the consequential photochemical impact of extended H$_2$O cross sections.

*Unified Astronomy Thesaurus concepts:* Planetary atmospheres (1244); Habitable planets (695); Exoplanet atmospheres (487); Exoplanets (498); Carbon dioxide (196)

## 1. Introduction

The search for life outside our solar system is centered around planets like Earth: small, rocky planets with secondary atmospheres (L. Kaltenegger 2017; E. W. Schwieterman et al. 2018). In the current state of exoplanet science, JWST represents a prime opportunity to observe and characterize the atmospheres of these Earth-sized terrestrial exoplanets that orbit in the habitable zones of their host stars (TRAPPIST-1 JWST Community Initiative et al. 2024; E. M. R. Kempton & H. A. Knutson 2024). However the most compelling targets for observations by JWST are those planets orbiting M-type stars (C. V. Morley et al. 2017; E. M. May et al. 2023). In 2021, the Astronomy & Astrophysics Decadal Survey, which highlights the scientific priorities, opportunities, and funding recommendations for the next decade, listed identifying and characterizing terrestrial exoplanets as a key goal (National Academies of Sciences, Engineering, and Medicine 2021). With this, one of the survey's top priorities is the development of the Habitable Worlds Observatory (HWO). HWO will be optimized for observing reflected light from small planets orbiting Sun-like host stars in the IR, optical, and UV (E. Mamajek & K. Stapelfeldt 2024).

We can use photochemical modeling to predict the possible atmospheres of some of the many exoplanets that have been discovered to date (J. Krissansen-Totton et al. 2018; A. P. Lincowski et al. 2018; R. Hu et al. 2021; N. Madhusudhan et al. 2023; V. S. Meadows et al. 2023). Forward modeling is also critical in helping to inform the development and design specifications of future exoplanet observing missions, as well as in interpreting observed exoplanet spectra (T. P. Greene et al. 2016; M. H. Currie et al. 2023; N. F. Wogan et al. 2024). The models used in photochemical studies require accurate inputs, including chemical reaction rates, molecular absorption cross sections, stellar spectra, dry and wet deposition rates, mixing parameterizations, and more. Models that are set with specific conditions in mind (often for the Earth or other solar system worlds) can err in their predictions when they are used to model atmospheres with substantially different boundary conditions. Additionally, models that combine incompatible photochemical inputs can yield erroneous results, and can lead to conflicting interpretations of observations.

Previous studies have shown that updates to the H$_2$O mid-UV (MUV; 200–300 nm) absorption cross sections can meaningfully impact predictions of trace gas chemistry on anoxic, temperate, terrestrial exoplanets (S. Ranjan et al. 2020; W. Broussard et al. 2024). Past H$_2$O cross-section prescriptions cut off at ∼200–208 nm, where H$_2$O's opacity falls below the typical scattering opacity of the atmosphere. However, in thick anoxic atmospheres, the 200–240 nm range is critical for the atmospheric chemistry (J.-S. Wen et al. 1989). This is because stellar MUV photons penetrate further into the H$_2$O-rich troposphere, whereas the higher-energy far-UV (<200 nm) photons are stopped from reaching the troposphere by overlying $CO_2$. With newly measured $H_2O$ cross sections that extend into the MUV, more $H_2O$ photolysis following Equation (1) will occur, generating more of the highly reactive OH radical.

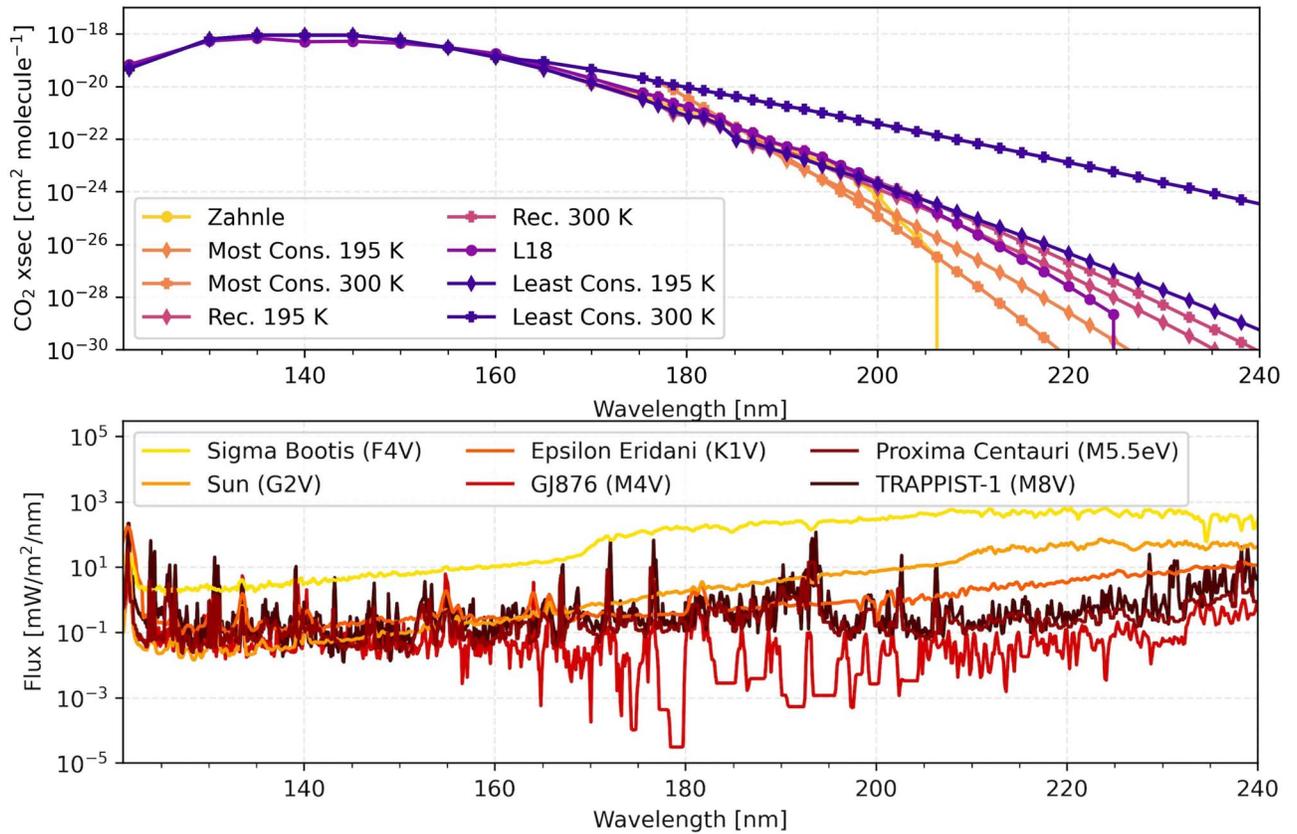

**Figure 1.** Top: $CO_2$ cross-section prescriptions from 120 to 240 nm. Bottom: spectral energy distribution of the stars modeled in this study, from 120 to 240 nm.

$$H_2O + h\nu \rightarrow H + OH \quad (1)$$

OH is an effective sink for many atmospheric trace gases, such as CO and $CH_4$. With more OH available to remove these gases, the extended $H_2O$ cross sections lead to lower predicted volume mixing ratios of atmospheric trace gases for these species.

There are two primary channels for the photolysis of $CO_2$:

$$CO_2 + h\nu \rightarrow CO(^1\Sigma^+) + O(^1D) \quad (2)$$

$$CO_2 + h\nu \rightarrow CO(^1\Sigma^+) + O(^3P). \quad (3)$$

The channel in Equation (2) has a quantum limit at 167.2 nm while the channel in Equation (3) has a quantum limit of 227.5 nm, representing the energy needed to break the CO–O bond in the ground state (J. A. Schmidt et al. 2013). Beyond this quantum limit, only forbidden transitions can occur, but the accumulation of these forbidden transitions can still potentially add to the opacity of the cross sections. For this reason, as seen in Figure 1, our extrapolation prescriptions continue beyond 227.5 nm, to account for this cumulative effect. Termination wavelengths for $CO_2$ cross sections vary from model to model. Most databases currently recommend a cutoff near ~200 nm (S. P. Sander et al. 2011), which was used by both S. Ranjan et al. (2020) and W. Broussard et al. (2024) and lies predictably near the wavelength where scattering opacities begin to overwhelm dissociation opacities (D. Ityaksov et al. 2008). A. P. Lincowski et al. (2018) used a log-extrapolated cutoff at 225 nm when modeling the photochemistry of the TRAPPIST-1 planets. The public version of the Atmos photochemical model (G. Arney et al. 2016) uses a cutoff prescription of ~208 nm, between these end-members.

$CO_2$ inputs are particularly consequential, as our definition of a traditionally habitable, Earth-like world is predicated on an $N_2$–$CO_2$–$H_2O$ atmosphere with a negative carbon feedback cycle (R. K. Kopparapu et al. 2013). Such worlds are also expected as a consequence of planetary outgassing (F. Gaillard & B. Scaillet 2014). The results of W. Broussard et al. (2024) showing the impact of the extended $H_2O$ cross sections into the MUV were based on simulations that did not include $CO_2$ absorption in this wavelength region. If $CO_2$ has appreciable absorption in the MUV, MUV photons will be stopped from reaching the $H_2O$-rich troposphere by the overlying $CO_2$, thus the impact of the extended $H_2O$ cross sections could be minimized. Likewise the introduction of CO, O($^3P$), and O($^1D$) from the photolysis of $CO_2$ can consume OH radicals produced via $H_2O$ photolysis, further contributing to the suppression of OH and minimizing the impact of the $H_2O$ cross sections. Furthermore, the photochemical production of CO, O($^3P$), and O($^1D$) has a variety of other spectral and chemical implications, including those assessing the potential for abiotic $O_2$ and $O_3$ accumulation and resulting spectral signatures that could constitute a false positive for life (e.g., P. Gao et al. 2015; C. Harman et al. 2015; E. W. Schwieterman et al. 2016; S. Ranjan et al. 2020). It is therefore essential to understand the degree to which corresponding extended $CO_2$ cross sections would impact those earlier results. In the absence of laboratory measurements of $CO_2$'s cross sections at room temperature in the MUV, we can use extrapolations to predict these data. In this paper, we use three cross-section extrapolations (described further in Section 2.1) in addition to the prescription used by A. P. Lincowski et al. (2018; the "L18" prescription), and the prescription used in the public version of Atmos (the "Zahnle" prescription; G. Arney et al. 2016). The three extrapolations, shown in Figure 1, include a most conservative extrapolation (which has the least opacity and thus leads to the least $CO_2$ photolysis), a least conservative extrapolation (which has the most opacity, leading to the most $CO_2$ photolysis), and a recommended extrapolation (which lies between these end-members).

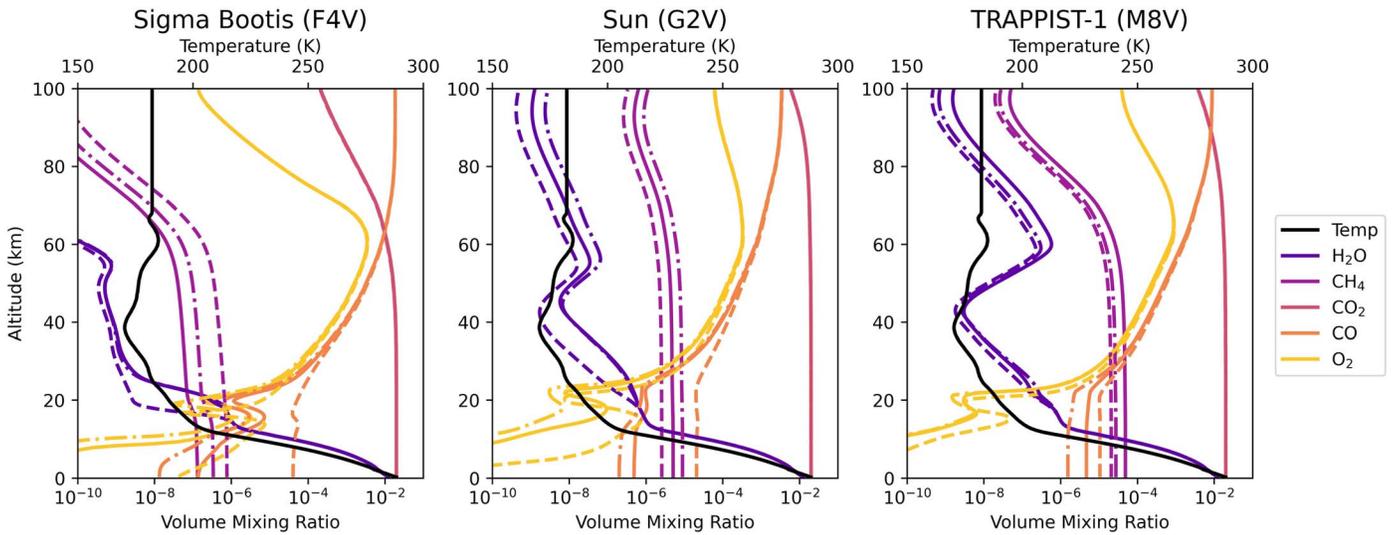

**Figure 2.** Example profile plots for planets orbiting Sigma Boötis, the Sun, and TRAPPIST-1, for a $CH_4$ flux of $2 \times 10^{10}$ molecules $cm^{-2}$ $s^{-1}$ and a $CO_2$ volume mixing ratio of 3%. Solid lines show the altitude-dependent volume mixing ratios of key atmospheric gases modeled using the recommended $CO_2$ cross sections; dashed lines are modeled using the least conservative $CO_2$ cross sections, and dashed–dotted lines are modeled using the most conservative $CO_2$ cross sections.

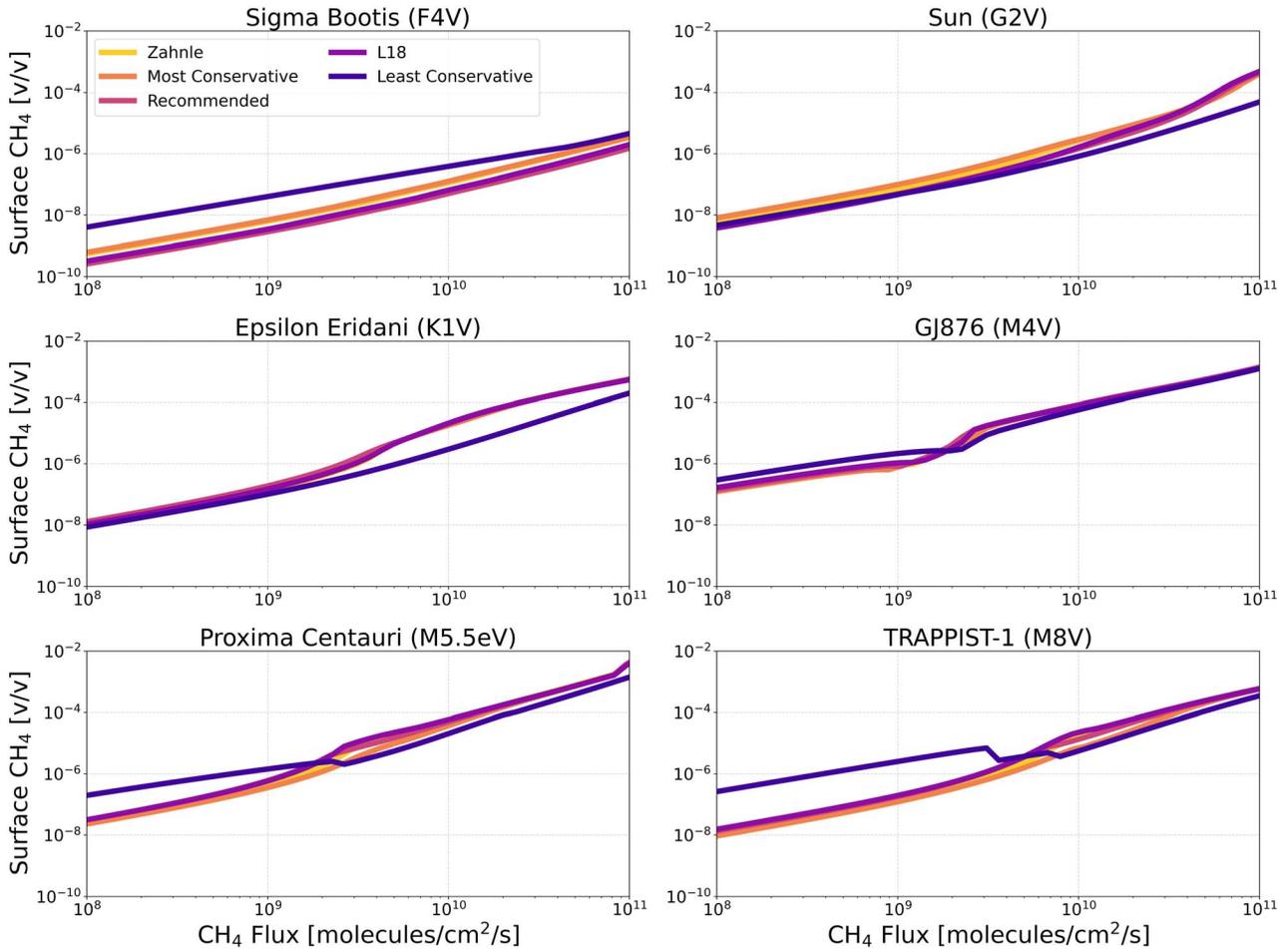

**Figure 3.** $CO_2$ cross-section sensitivity test: surface $CH_4$ vs. $CH_4$ flux for anoxic habitable planets orbiting FGKM-type host stars.

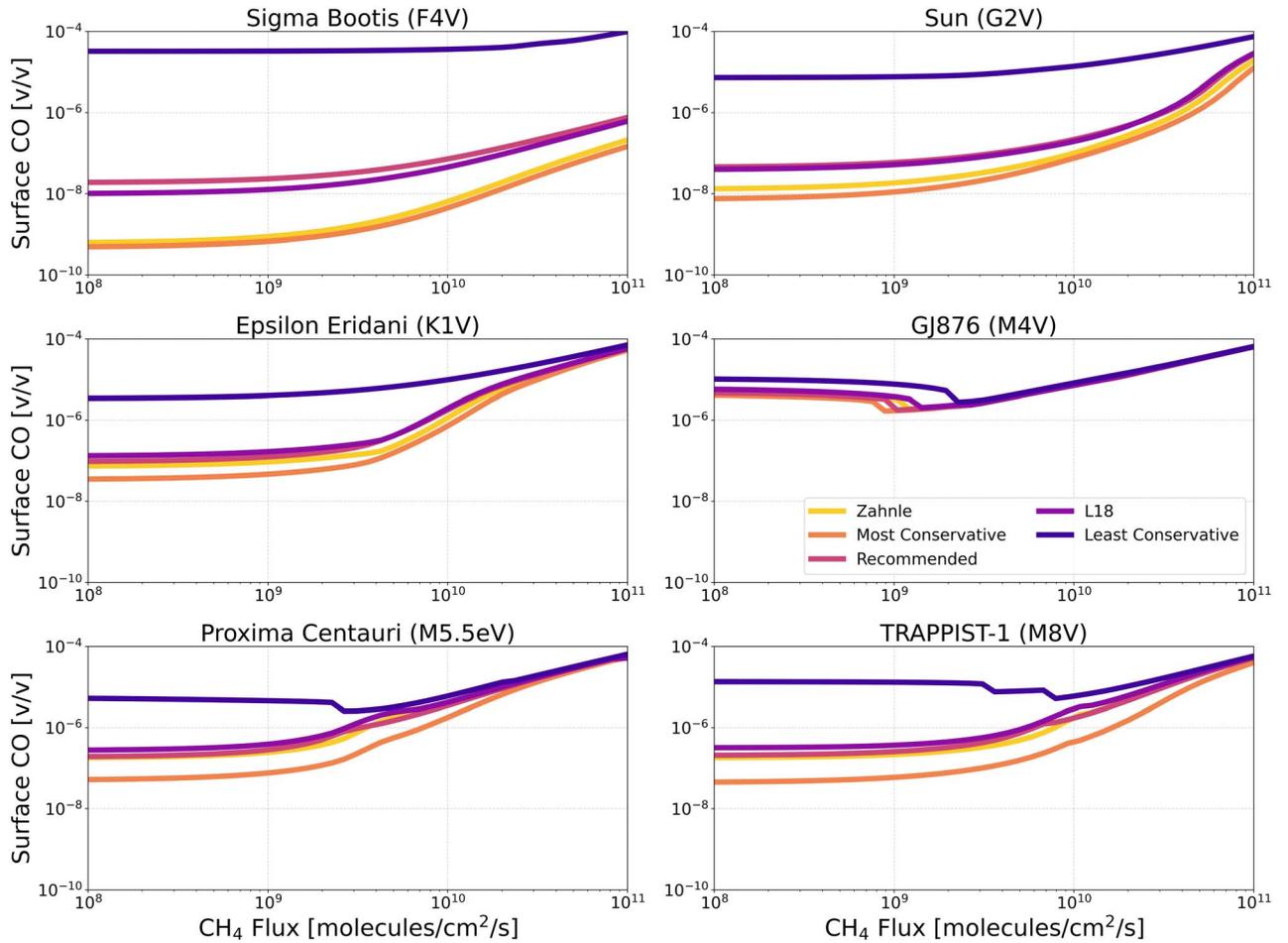

**Figure 4.** $CO_2$ cross-section sensitivity test: surface CO vs. $CH_4$ flux for anoxic habitable planets orbiting FGKM-type host stars.

**Table 1**
$CO_2$ Cross-section Experimental Data Used for Extrapolations

| Temperature (K) | Wavelength Range (nm) | Extrapolation | Data Source |
|---|---|---|---|
| 195 | 163–192.5 | recommended | O. Venot et al. (2018) |
| 195 | 163–192.5 | most conservative | W. Parkinson et al. (2003) |
| 195 | 163–192.5 | least conservative | W. Parkinson et al. (2003) |
| 295 | 163–200 | recommended | W. Parkinson et al. (2003) |
| 300 | 115.3–187.5 | most conservative | O. Venot et al. (2018) |
| 300 | 115–200 | least conservative | O. Venot et al. (2018) |

In this paper, we test the impact of these prescriptions for $CO_2$'s cross sections in temperate, anoxic, terrestrial planet atmospheres, and make recommendations for harmonizing the extended $CO_2$ cross-section inputs in community models. In Section 2, we describe the construction of the extended $CO_2$ cross sections via empirical and theoretical sources (a "best-fit" range), comparing them quantitatively to current prescriptions. Additionally we describe the photochemical and spectral models, as well as the planetary scenario used to test the sensitivity of these inputs under a range of $CH_4$ surface fluxes and $CO_2$ surface volume mixing ratios, for FGKM-type host stars. In Section 3 we report our results, including impacts on trace gas species and spectral observables. We also revisit the $H_2O$ cross-section sensitivity tests of W. Broussard et al. (2024) with these updated $CO_2$ cross sections. We discuss the implications of our results in Section 4 and conclude in Section 5.

## 2. Methods

### 2.1. Cross-section Prescriptions

The $CO_2$ cross sections were prepared using a similar prescription to the extrapolation of the $H_2O$ cross sections presented by S. Ranjan et al. (2020) and used by W. Broussard et al. (2024). Under a first-order assumption of a linear-log loss of opacity toward dissociation, over large wavenumber bins, several possible extrapolations were proposed based on existing experimental data measured primarily at wavelengths ⩽200 nm (Table 1). Our extrapolations were created by

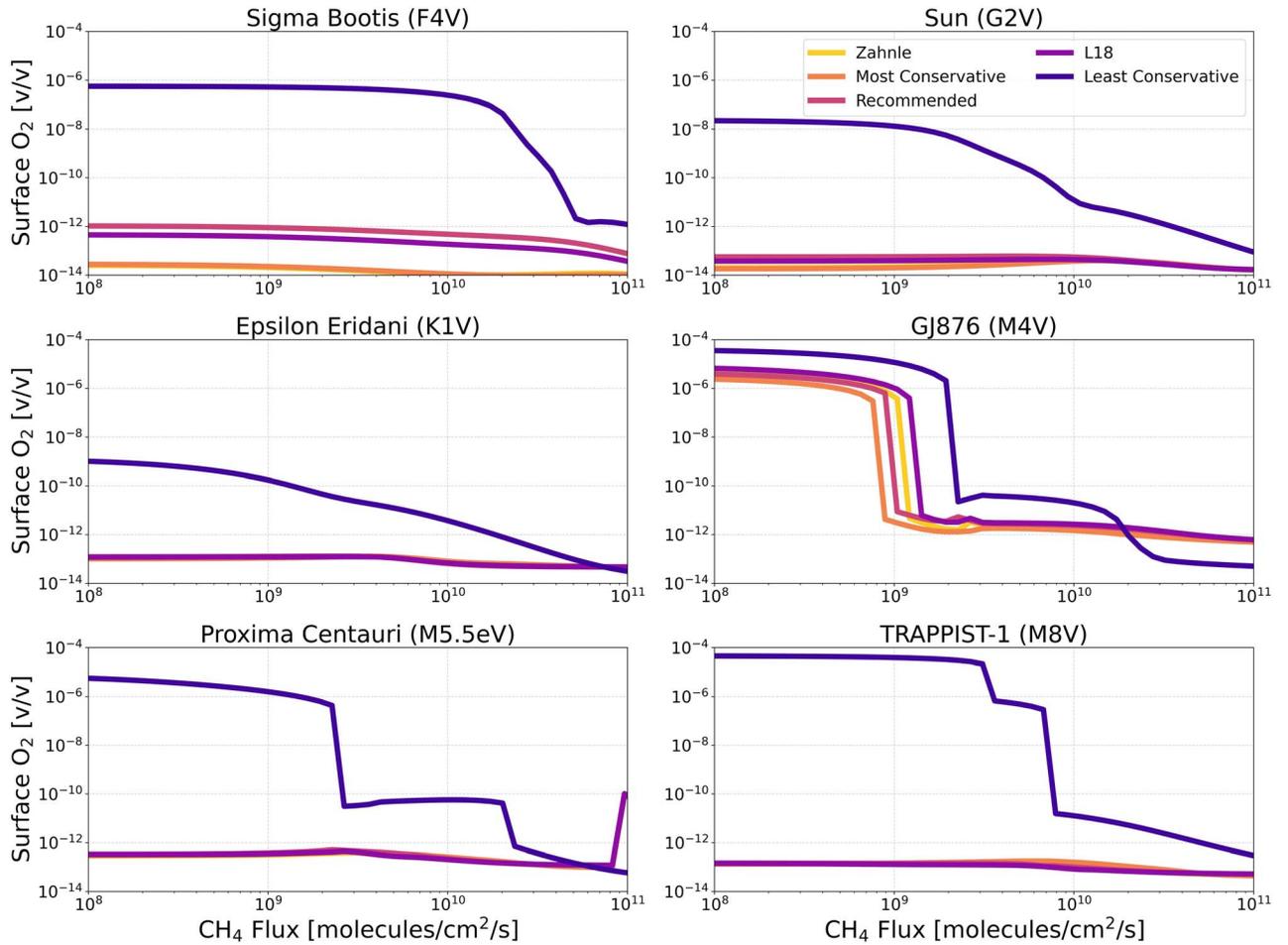

**Figure 5.** $CO_2$ cross-section sensitivity test: surface $O_2$ vs. $CH_4$ flux for anoxic habitable planets orbiting FGKM-type host stars.

**Table 2**
Stellar and Planetary Properties

| Star | Spectral Type | $T_{\rm eff}$ (K) | Luminosity ($L_\odot$) | Stellar Radius ($R_\odot$) | Distance (pc) | Planetary Radius ($R_\oplus$) | Planet–Star Distance (au) |
|---|---|---|---|---|---|---|---|
| σ Boötis | F4V | 6435 | 3.1541 | 1.4307 | 15.8 | 1 | 1.776 |
| Sun | G2V | 5780 | 1 | 1 | … | 1 | 1 |
| ε Eridani | K1V | 5039 | 0.32 | 0.735 | 3.2 | 1 | 0.562 |
| GJ 876 | M4V | 3129 | 0.0122 | 0.3761 | 4.69 | 1 | 0.110 |
| Proxima Centauri | M5.5eV | 2992 | 0.001567 | 0.147 | 1.3 | 1 | 0.049 |
| TRAPPIST-1 | M8V | 2559 | 0.000524 | 0.117 | 12.1 | 0.91 | 0.029 |

following the gradient, calculated from

$$(\log(\sigma_{\lambda_b}) - \log(\sigma_{\lambda_a}))/\log(\sigma_{\lambda_b}) \quad (4)$$

from the baseline of the clear rovibrational structures present in the measured data, where $\sigma$ is the absorption cross section. We selected structures that were toward the end of the measured data, and therefore toward the end of the instrument sensitivity, but otherwise at a long wavelength so as to create a more representative logarithmic trend. The range of opacities in the predicted cross sections come primarily from the variation in accuracy and precision of the measured data used as a basis for the theoretical extrapolations. A secondary reason for the variation in extrapolations derives from the range of possible linear-log gradients that can be predicted from the overall opacity-loss trend in the rovibrational structure of the measured data. For every presented temperature, three cross-section extrapolations were created (recommended, and most/least conservative) to allow for a scientifically meaningful sensitivity analysis of the impact of these cross sections on the photochemical models explored in this work. Note that "least conservative" extrapolations have the highest dissociation opacity at wavelengths ⩾200 nm, while the "most conservative" extrapolations have the lowest dissociation opacity at wavelengths ⩾200 nm. Notably, due to the higher complexity of rovibrational features measured at low temperatures, the extrapolated linear-log gradients did not always decrease with lower temperatures, despite there being no physical reason to expect higher opacities at lower temperatures. For example, the most conservative extrapolation at 300 K is well below all of the extrapolations at 195 K as seen in Figure 1. However, not all of the extrapolations show this artifact, as other higher-temperature extrapolations exhibit higher opacities as expected. Nonetheless, this exception only applies to the extreme (most and least conservative) extrapolations, not to the recommended opacities.

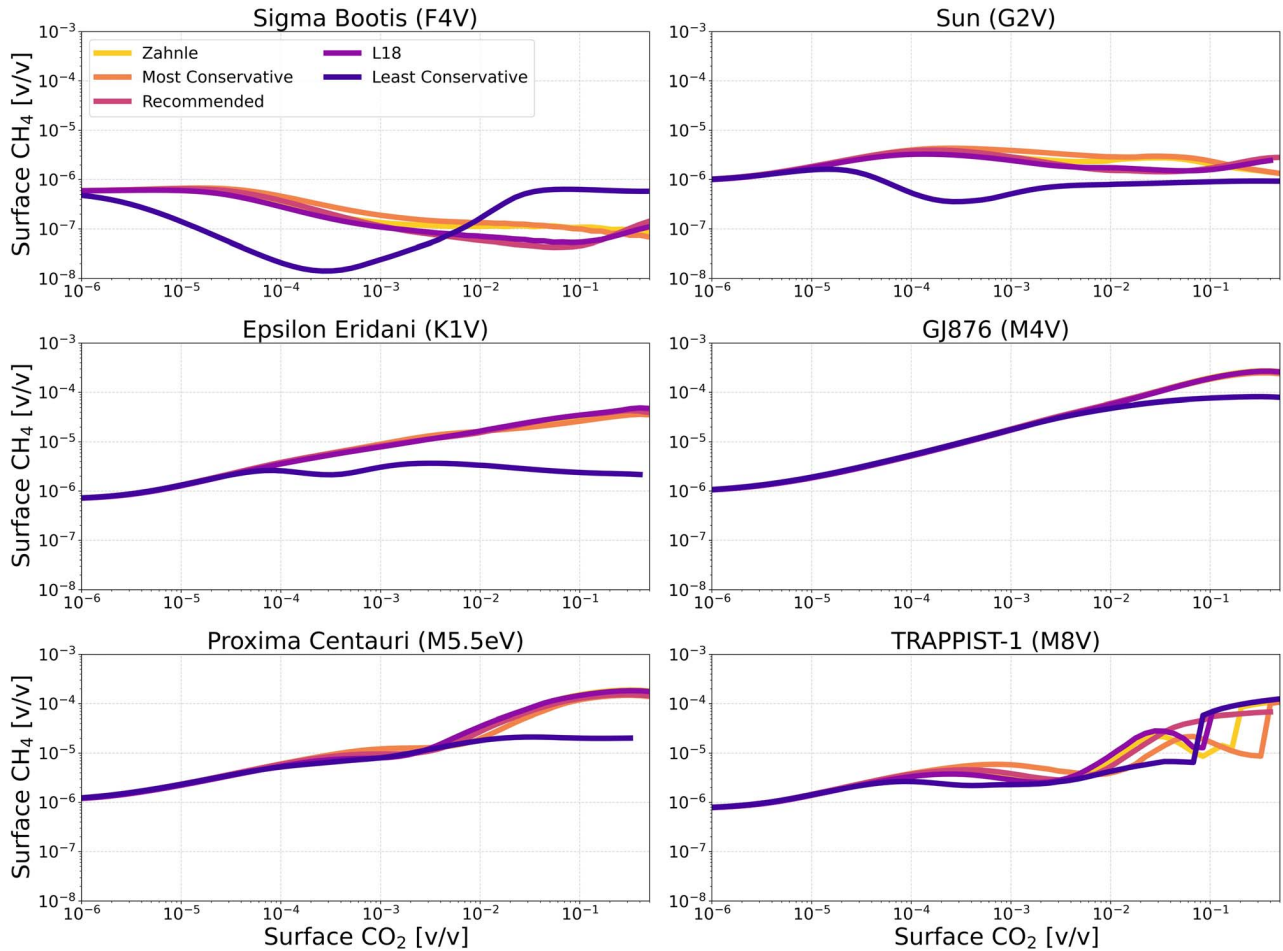

**Figure 6.** $CO_2$ cross-section sensitivity test: surface $CH_4$ vs. surface $CO_2$ for anoxic habitable planets orbiting FGKM-type host stars.

It is worth noting that the empirically calibrated but theoretical $CO_2$ cross sections used in this work should be considered only as a reasonable substitution given the absence of data at those wavelengths—valuable for sensitivity analyses, but not an adequate replacement for accurate measured data, which we recommend should be funded and obtained. It is reasonable to assume, however, that the true cross sections, once measured, will fall in the range of theoretical extrapolations measured here.

Table 1 lists the sources used to create the theoretical extrapolations of the $CO_2$ cross sections. The cross-section extrapolations are available at https://github.com/abrou009/broussard_2025_co2.

## 2.2. Photochemical Model Description

This research employs the photochemical model Atmos (G. Arney et al. 2016, 2018; R. C. Felton et al. 2022; E. W. Schwieterman et al. 2022), the same one-dimensional model as was used by W. Broussard et al. (2024). For a more detailed description of the photochemical portion of Atmos, see Section 2.2 of W. Broussard et al. (2024). Also contained within Atmos is the radiative–convective climate model, Clima. Clima was first developed to model high $CO_2$ concentrations in the early Earth's atmosphere (J. F. Kasting & T. P. Ackerman 1986), but has since received updates to model a broader range of climate scenarios (G. Arney et al. 2016). Clima can be run independently of the photochemistry portion of Atmos, or both portions can be run in the coupled mode. To use Atmos in the photochemistry–climate coupling mode, first the photochemical portion is run to convergence. The altitude-dependent volume mixing ratios of relevant gas species are returned once convergence is reached, and this output is used as the initial input into the climate portion of the model. Clima updates the water vapor and temperature profiles, which are then returned to the photochemistry portion of Atmos, and this cycle repeats until both the photochemistry and climate models have converged.

In this research, our utilization of Atmos differs from that of W. Broussard et al. (2024) in two ways: in the chosen temperature–pressure profile and in the implementation of temperature-dependent cross sections for $CO_2$. The temperature–pressure profile used in this research has a surface temperature of 288 K (Earth's average modern surface temperature) and was calculated to be climatically self-consistent up to 68.25 km, which represents the model top of the climate simulations. Above this height the atmosphere takes on an isothermal profile with a temperature of 182 K. This profile represents a reasonable approximation of a temperature–pressure profile for the Archean Earth, and is used for each of the stars modeled in this study to isolate the effects of the changing $CO_2$ cross sections and one other variable (e.g., $CH_4$ production rate, $CO_2$ mixing ratio, stellar spectrum, etc.). For all planets except those orbiting Proxima Centauri and TRAPPIST-1, the stellar spectra were scaled so that the planet received a top-of-atmosphere flux equal to the solar constant. We note that past climate modeling has shown that planets that receive an Earth-average insolation flux will have a different surface temperature depending on the spectral energy distribution of the host star (A. A. Segura et al. 2005; G. Arney et al. 2018; A. D. Del Genio et al. 2019). M dwarf host stars produce more red and infrared light than stars of earlier types, which is more easily transmitted (not scattered) through a planetary atmosphere. Consequently, planets orbiting M-type host stars will have higher surface temperatures for a given bolometric flux than planets orbiting G- or F-type host stars, where proportionally more of the radiation received at the top of the atmosphere is scattered away, increasing planetary albedo. As the changing tropospheric water content resulting from the changing surface temperature would have a strong impact on our results, we use the same surface temperature for all planets regardless of the stellar type of the host star as a simplifying assumption, allowing us to facilitate direct intercomparisons between the host stars and $CO_2$ cross sections.

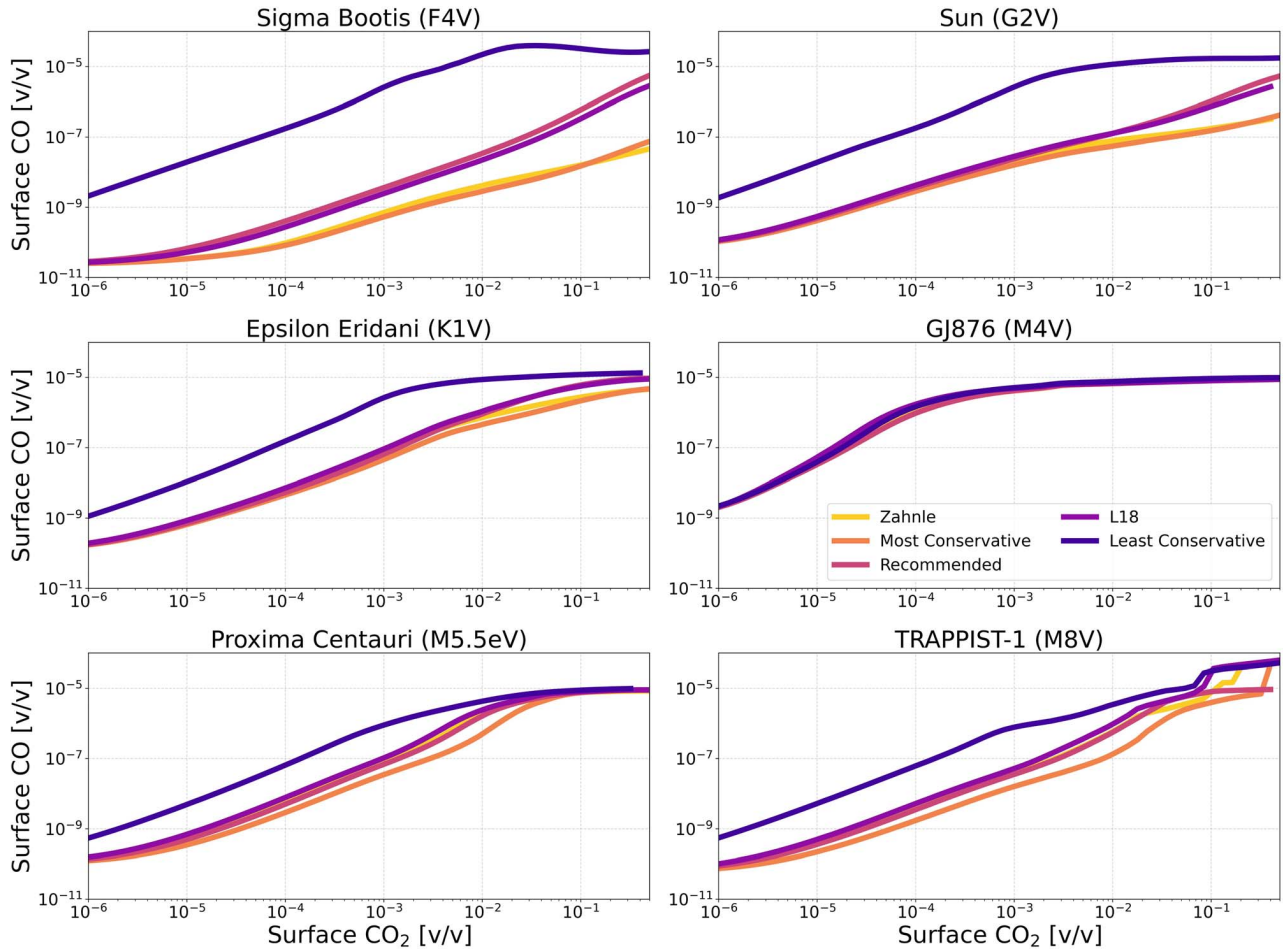

**Figure 7.** $CO_2$ cross-section sensitivity test: surface CO vs. surface $CO_2$ for anoxic habitable planets orbiting FGKM-type host stars.

We can factor in some of the temperature-dependent nature of the photochemical cross sections, and account for the majority of the altitude-dependent temperature variation anticipated for a habitable planet's atmosphere, by including an interpolation between the prescriptions at each provided temperature. Thus, we use a temperature-dependent linear interpolation between the 195 K and 300 K cross-section data sets. For temperatures greater than 300 K, the 300 K cross sections would be assumed, and the 195 K cross sections would be assumed for temperatures less than 195 K.

### 2.3. Spectral Model Description

This research employs the same spectral model as W. Broussard et al. (2024), the Spectral Mapping Atmospheric Radiative Transfer code (SMART; V. S. Meadows & D. Crisp 1996; D. Crisp 1997). For a more detailed description of SMART, see Section 2.3 of W. Broussard et al. (2024). For the purpose of this research, we have modeled spectral scenarios assuming Sigma Boötis, the Sun, and TRAPPIST-1 as the host star. We give the stellar and planetary parameters assumed in Table 2. Planetary parameters were chosen for Sigma Boötis, Epislon Eridani, and GJ 876 as a host star so that the planet would have a solar constant equal to the Earth's current solar constant. For Proxima Centauri and TRAPPIST-1 as the host stars, the planetary parameters of Proxima Centauri b and TRAPPIST-1 e were assumed, respectively.

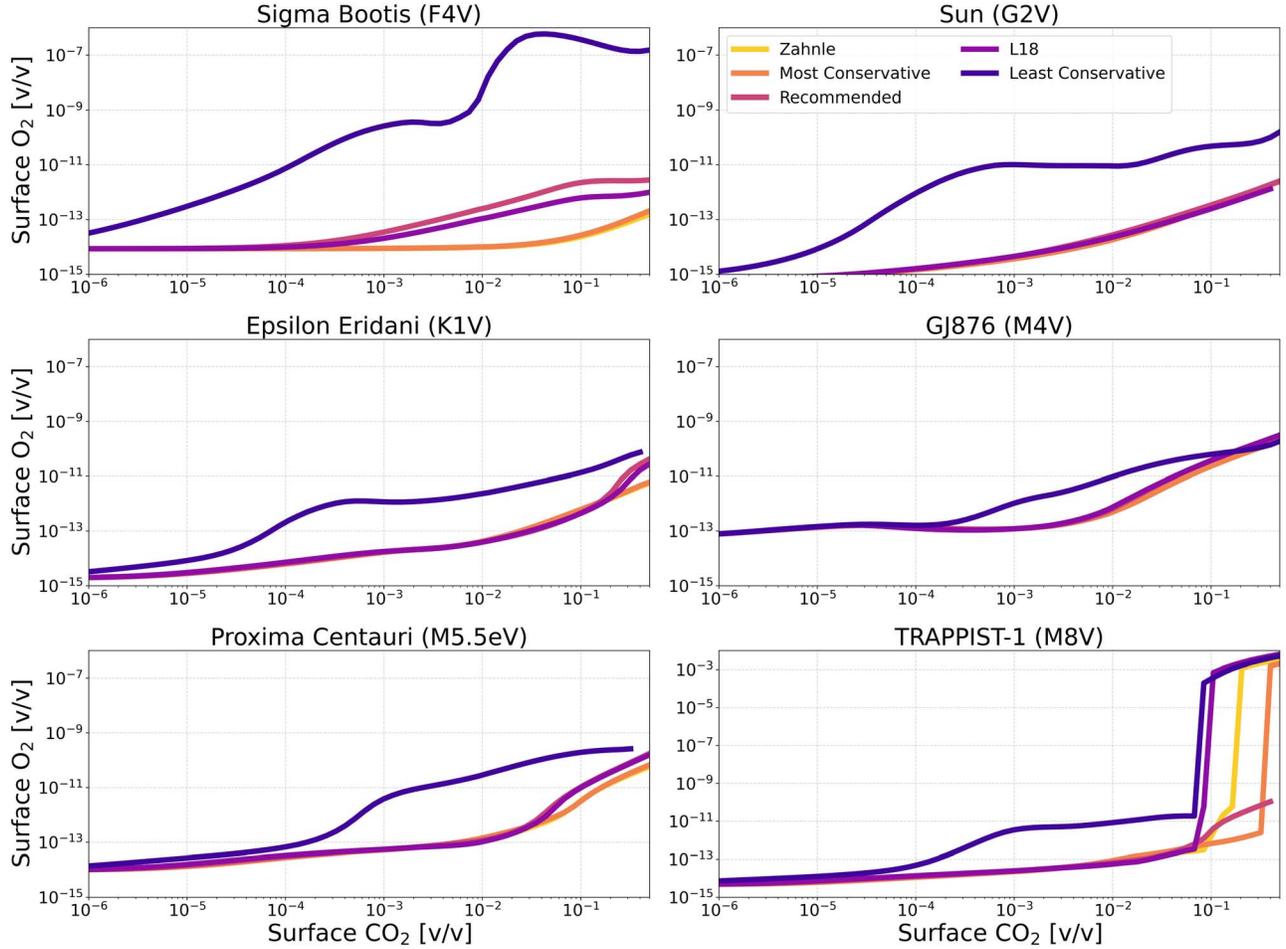

**Figure 8.** $CO_2$ cross-section sensitivity test: surface $O_2$ vs. surface $CO_2$ for anoxic habitable planets orbiting FGKM-type host stars.

### 2.4. Stellar Spectra

To show how our results vary with the type of host star, we chose to conduct these sensitivity tests with six different main-sequence host stars, with stellar parameters listed in Table 2. These stars are: the F-type star $\sigma$ Boötis (A. Segura et al. 2003), the Sun, which is a G-type star (G. Thuillier et al. 2004), the K-type star $\epsilon$ Eridani (A. Segura et al. 2003), and three M-type stars: GJ 876 (M4V) (K. France et al. 2016; A. Youngblood et al. 2016; R. O. P. Loyd et al. 2016 (v22)), Proxima Centauri (M5.5eV) (E. L. Shkolnik & T. S. Barman 2014; R. O. P. Loyd et al. 2018; S. Peacock et al. 2020), and TRAPPIST-1 (M8V) (S. Peacock et al. 2019a, 2019b). Figure 1 shows the spectral energy distribution of each of these host stars in the UV.

### 2.5. Planetary Scenario

The atmospheres modeled in this work are $N_2$–$H_2O$–$CO_2$ atmospheres. Full atmospheric boundary conditions, including surface fluxes, surface volume mixing ratios, and dry deposition velocities, can be found in Table 3 of Appendix A. Figure 2 shows three example profile plots for (from left to right) Sigma Boötis, the Sun, and TRAPPIST-1 as the host star, for a surface $CH_4$ flux of $2 \times 10^{10}$ molecules cm$^{-2}$ s$^{-1}$ and a $CO_2$ volume mixing ratio of 3%.

For conducting the $CH_4$ flux sensitivity tests, we adopt a $CO_2$ mixing ratio of 3% and vary the $CH_4$ surface flux from $10^8$ molecules cm$^{-2}$ s$^{-1}$ ($\sim 2.67 \times 10^{-2}$ Tmol yr$^{-1}$), which is around the expected $CH_4$ flux values for abiotic systems, such as from volcanic outgassing or serpentinization (M. A. Thompson et al. 2022), to $10^{11}$ molecules cm$^{-2}$ s$^{-1}$ ($\sim 26.7$ Tmol yr$^{-1}$), representing a roughly Earth-like flux; Earth's current $CH_4$ production levels are around 30 Tmol yr$^{-1}$ (M. A. Thompson et al. 2022).

For the $CO_2$ mixing ratio sensitivity tests, we adopt a $CH_4$ flux of $10^9$ molecules cm$^{-2}$ s$^{-1}$ and vary the $CO_2$ surface volume mixing ratio from $10^{-6}$, or 1 ppm, to $\sim 5 \times 10^{-1}$, or about 50% $CO_2$.

## 3. Results

### 3.1. Relationships between $CO_2$ Cross Sections and Trace Gases

To test the sensitivity of atmospheric trace gases to the choice of $CO_2$ cross-section prescription, we have conducted two sets of sensitivity tests. The first set, described in Section 3.1.1, test the response of atmospheric trace gases to $CH_4$ surface flux, going from a $CH_4$ flux of $10^8$ molecules cm$^{-2}$ s$^{-1}$ to $10^{11}$ molecules cm$^{-2}$ s$^{-1}$ (or from $\sim 2.67 \times 10^{-2}$ Tmol yr$^{-1}$ to $\sim 26.7$ Tmol yr$^{-1}$). As in W. Broussard et al. (2024), this parameter space was chosen to represent a gradient that goes from an abiotic value less than the upper limit of $CH_4$ from serpentinization to a biotic level of $CH_4$, similar to Earth's current biogenic $CH_4$ flux (M. A. Thompson et al. 2022). The second set of sensitivity tests, described in Section 3.1.2, test the response of atmospheric trace gases as a function of the surface $CO_2$ volume mixing ratio, varying from $10^{-6}$ to around $5 \times 10^{-1}$ (or from 1 ppm to around 50% $CO_2$).

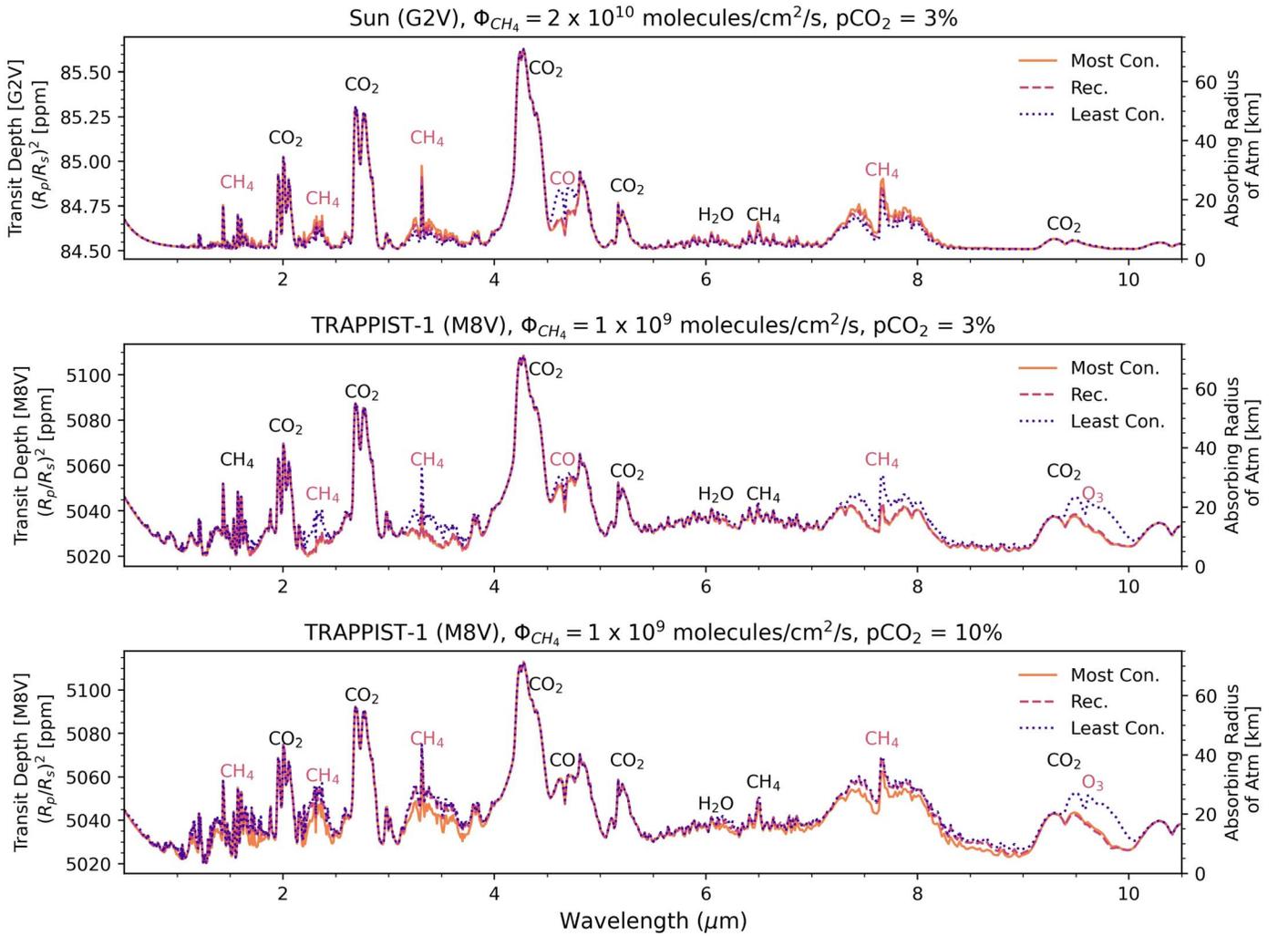

**Figure 9.** Comparison of transmission spectra between the recommended, most conservative, and least conservative $CO_2$ prescriptions, for three different scenarios. The top panel shows transmission spectra for a planet orbiting the Sun, with a $CH_4$ flux = $2 \times 10^{10}$ molecules cm$^{-2}$ s$^{-1}$, with a surface $CO_2$ volume mixing ratio of 3%. The middle and bottom panels show the transmission spectra for a planet orbiting TRAPPIST-1 with a $CH_4$ flux = $1 \times 10^9$ molecules cm$^{-2}$ s$^{-1}$; the middle also has a surface $CO_2$ volume mixing ratio of 3%, while the bottom panel has a surface $CO_2$ volume mixing ratio of 10%. Features that are shared between spectra are labeled in black; features that differ depending on the cross-section prescription are labeled in red.

### 3.1.1. $CH_4$ Flux Sensitivity Tests

Figures 3, 4, and 5 show the impact of the various $CO_2$ cross-section prescriptions on the surface volume mixing ratios of $CH_4$, CO, and $O_2$ as a function of $CH_4$ surface flux, respectively.

In Figure 3, we see that the choice of $CO_2$ cross-section prescription does not have a large impact on the resulting surface $CH_4$ abundance. Four of the five cross-section prescriptions result in almost identical model predictions across the six stellar types, with one exception in the results from the least conservative prescription (which has the largest $CO_2$ opacity at wavelengths >200 nm). For the F-type host star and the M5.5eV host star, at low $CH_4$ fluxes, the resulting surface $CH_4$ is around an order of magnitude larger than surface $CH_4$ modeled using the other prescriptions; for the M8V host star, the difference is a little over an order of magnitude. Additionally, for the G-type host star at the largest $CH_4$ fluxes, the surface $CH_4$ modeled using the least conservative prescription diverges from models run using the other prescriptions. Although there is less OH produced from $H_2O$ photolysis for models run using the least conservative prescription, there is also more $O(^3P)$ and $O(^1D)$ generated from the increased $CO_2$ photolysis. Thus, this increased $O(^3P)$ and $O(^1D)$ consume the $CH_4$, leading it to build up more slowly with the least conservative prescription.

As a direct product of $CO_2$ photolysis, the increased CO that is generated with the least conservative $CO_2$ cross-section prescription is readily apparent in Figure 4. In this figure we see large differences in the resulting surface CO abundance depending on the $CO_2$ cross-section prescription. Because of its high stellar flux overall, these differences are greatest for the

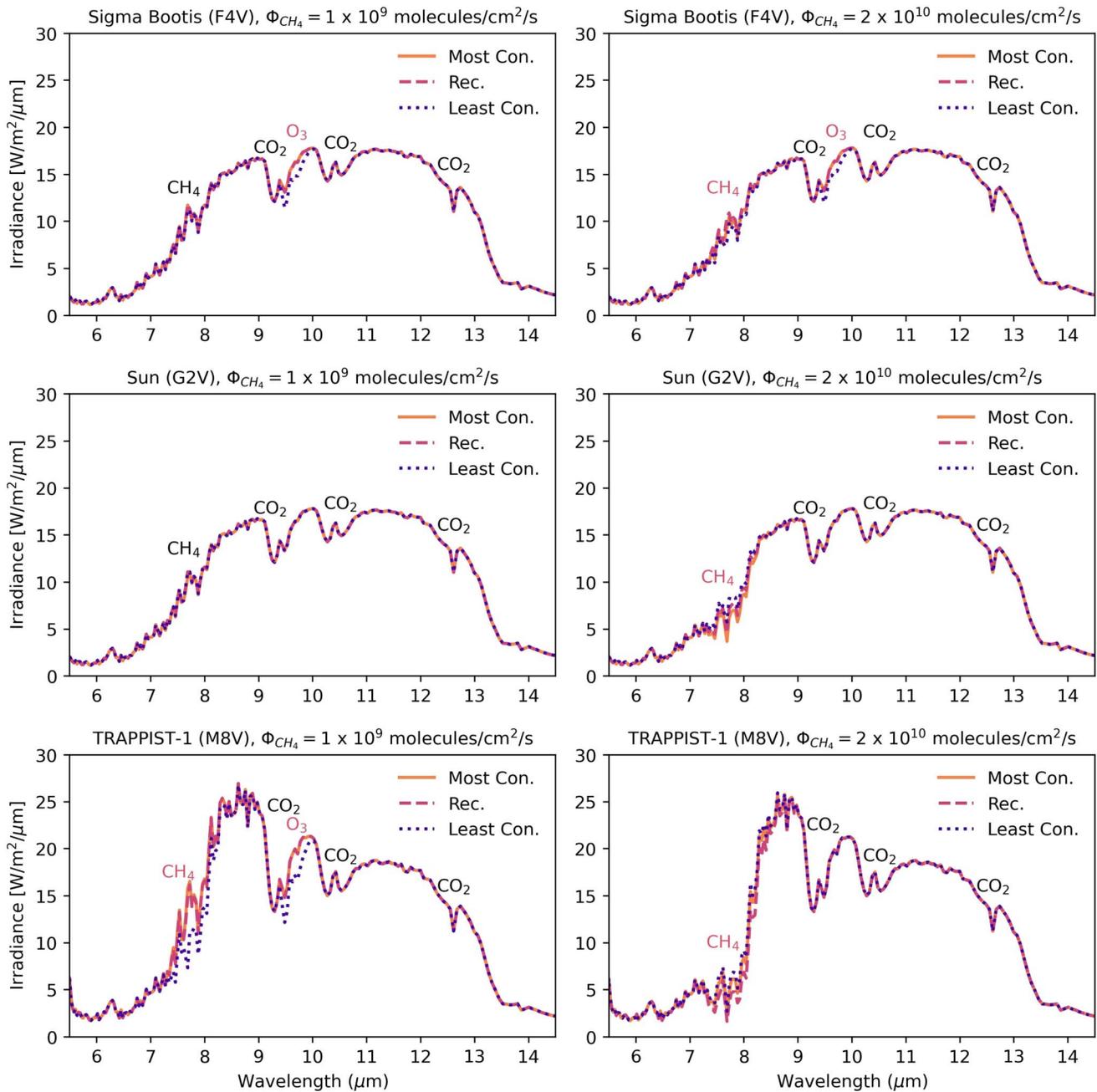

**Figure 10.** Comparison of emission spectra between the recommended, most conservative, and least conservative $CO_2$ prescriptions, for two different $CH_4$ fluxes (panels on the left are modeled with a $CH_4$ flux $= 1 \times 10^9$ molecules cm$^{-2}$ s$^{-1}$, those on the right with a $CH_4$ flux $= 2 \times 10^{10}$ molecules cm$^{-2}$ s$^{-1}$) with (from top to bottom) Sigma Boötis, the Sun, and TRAPPIST-1 as the host star. Features that are shared between spectra are labeled in black; features that differ depending on the cross-section prescription are labeled in red.

F-type host star at the lowest $CH_4$ flux, where there is over four orders of magnitude more surface CO modeled with the least conservative cross sections than with the most conservative.

The largest relative cross-section-dependent differences are seen with the surface $O_2$ volume mixing ratios, as seen in Figure 5. Here, we see that the source of variation in predicted surface $O_2$ is almost exclusive to the models run using the least conservative $CO_2$ cross sections, which predict over eight orders of magnitude more surface $O_2$ at low $CH_4$ fluxes for TRAPPIST-1 as the host star. There are two exceptions to this: the first, with Sigma Boötis as the host star, where there are almost two orders of magnitude difference between the recommended and most conservative prescriptions. However, this difference occurs at very low $O_2$ mixing ratios (e.g., $2.8 \times 10^{-14}$ predicted with the most conservative prescription, versus $1.0 \times 10^{-12}$ predicted with the recommended prescription), thus we would not expect a corresponding difference in the spectral observables of either $O_2$ or $O_3$. The other exception is with GJ 876 as the host star, which demonstrates a stepwise decrease in surface $O_2$ at a $CH_4$ flux of around $10^9$ molecules cm$^{-2}$ s$^{-1}$ regardless of the cross-section prescription. This is caused by variation in the threshold by which the $CH_4$ collapses the $O_2$ levels, and is particularly sensitive to input parameters such as the $CH_4$ flux and cross sections. This strong sensitivity of $O_2$ to reductant fluxes is commonly seen for late-type host stars, and is ultimately traced to their high far-UV/near-UV ratios (C. Harman et al. 2015; P. Barth et al. 2024).

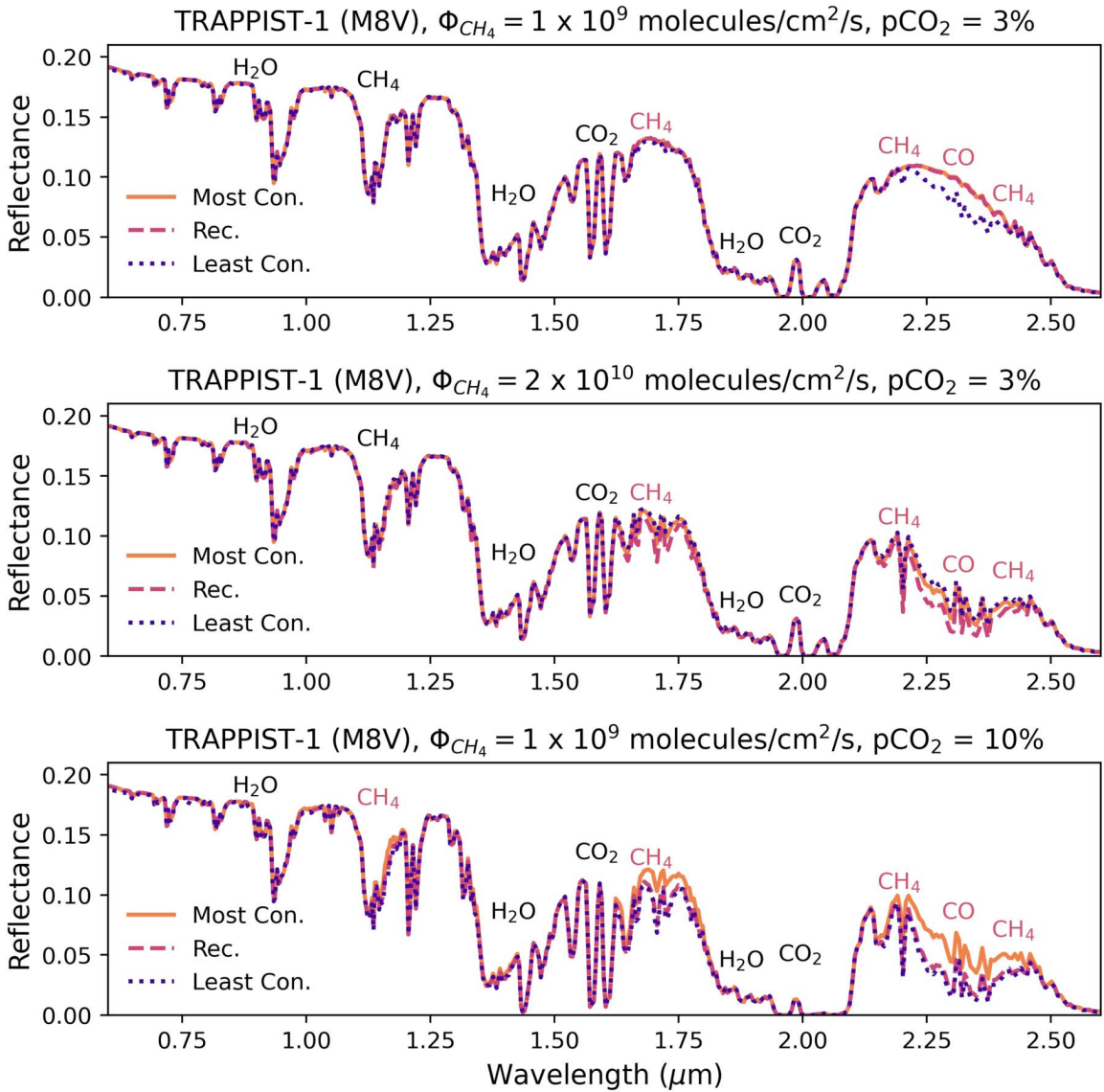

**Figure 11.** Comparison of reflection spectra between recommended, most conservative, and least conservative $CO_2$ prescriptions, for TRAPPIST-1 as a host star. The top panel is modeled with a $CH_4$ flux = $1 \times 10^9$ molecules cm$^{-2}$ s$^{-1}$ and a surface $CO_2$ volume mixing ratio of 3%, the middle panel is modeled with a $CH_4$ flux = $2 \times 10^{10}$ molecules cm$^{-2}$ s$^{-1}$ and a surface $CO_2$ volume mixing ratio of 3%, and the bottom panel is modeled with a $CH_4$ flux = $1 \times 10^9$ molecules cm$^{-2}$ s$^{-1}$ and a surface $CO_2$ volume mixing ratio of 10%. Features that are shared between spectra are labeled in black; features that differ depending on the cross-section prescription are labeled in red.

Appendix B shows an alternative visualization of these trace gas abundance differences at specific $CH_4$ fluxes of $10^8$, $10^9$, $10^{10}$, and $10^{11}$ molecules cm$^{-2}$ s$^{-1}$, for the F-, G-, and K-type host stars, as well as the M5.5eV host star in Figures 16, 17, and 18 in Appendix B.

*3.1.2. $CO_2$ Surface Mixing Ratio Sensitivity Tests*

Figures 6, 7, and 8 show the impact of the various $CO_2$ cross-section prescriptions on surface $CH_4$, CO, and $O_2$ as a function of $CO_2$ surface mixing ratio, respectively. Here, we can see that the predicted trace gas abundances do not vary greatly between $CO_2$ cross-section prescriptions, though where differences exist they are mostly exhibited by models run using the least conservative prescription.

Appendix B shows an alternative visualization of these trace gas abundance differences at specific $CO_2$ volume mixing ratios of fluxes of $10^{-6}$, $10^{-4}$, $10^{-2}$, and $3 \times 10^{-1}$, for the F-, G-, and K-type host stars, as well as the M5.5eV host star in Figures 19, 20, and 21 in Appendix B.

*3.2. Spectral Sensitivity*

Depending on the choice of $CO_2$ cross-section prescription, model predictions of atmospheric trace gas abundances can vary substantially, but that does not necessarily mean they will induce observable variations in planetary spectra. To demonstrate how these differing predictions may result in differences in the predicted observables, we model the transmission, emission, and reflected light spectra for various combinations of host star, $CH_4$ flux, and $CO_2$ mixing ratio, for models run using the most conservative, least conservative, and recommended $CO_2$ cross-section prescriptions. These spectra are shown in Figures 9, 10, and 11, respectively. Spectral features shared between spectra are labeled in black text, and spectral features that differ depending on the cross-section prescription are labeled in red text. In all scenarios, the spectra were generated with an assumed cloud coverage of 50% clear sky, 25% cirrus clouds, and 25% stratus clouds, as in G. Arney et al. (2016). Planetary parameters used to model the given scenarios are provided in Table 2.

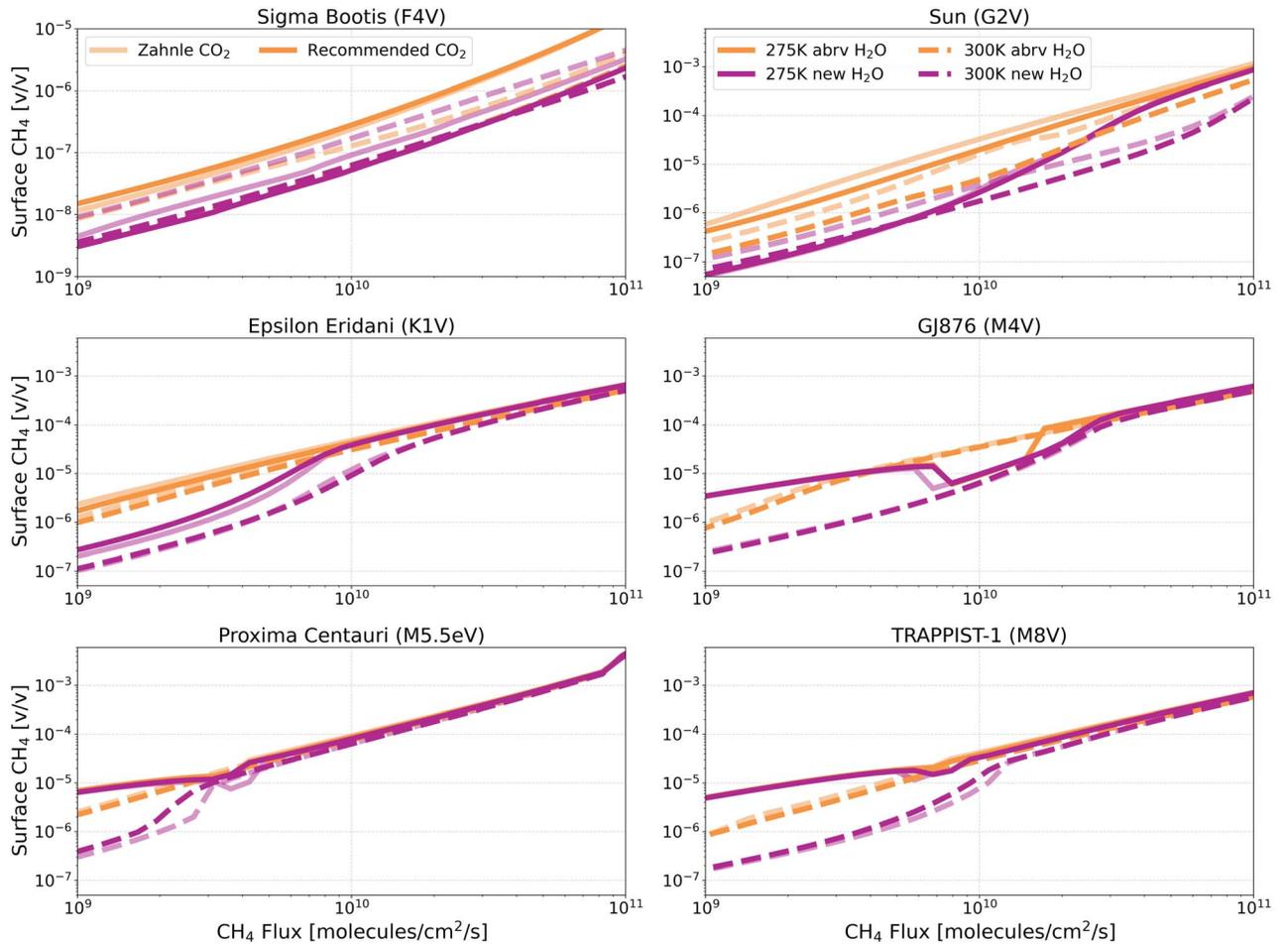

**Figure 12.** Comparison of $H_2O$ cross-section sensitivity tests with old and recommended $CO_2$ cross sections; surface $CH_4$ vs. $CH_4$ flux for anoxic habitable planets orbiting FGKM-type host stars.

In Figure 9, we show the transmission spectra for three different scenarios. The top panel is modeled using the Sun as a host star with a planetary $CH_4$ flux = $2 \times 10^{10}$ molecules $cm^{-2}$ $s^{-1}$ and an atmospheric $CO_2$ surface volume mixing ratio of 3%. This scenario corresponds to the example profile plot shown in the middle panel of Figure 2, and we can see that the greater levels of CO and lower levels of $CH_4$ modeled using the least conservative $CO_2$ cross sections lead to a larger CO feature around 4.6 $\mu$m and smaller $CH_4$ features around 1.7, 2.4, 3.3, and 7.6 $\mu$m. In the middle panel of Figure 9, we show transmission spectra modeled using TRAPPIST-1 as the host star, with planetary parameters emulating TRAPPIST-1 e, a $CH_4$ flux = $1 \times 10^9$ molecules $cm^{-2}$ $s^{-1}$ and a $CO_2$ surface volume mixing ratio = 3%. Here we see that the increased surface $CH_4$ predicted using the least conservative prescription leads to larger spectral features due to $CH_4$, particularly the $CH_4$ peak around 3.3 $\mu$m. In addition to the $CH_4$ features, the CO around 4.6 $\mu$m is slightly more prominent with the least conservative prescription, and, most notably, the spectrum modeled with the least conservative prescription shows an $O_3$ feature around 9.6 $\mu$m that is not seen in the other spectra of this panel. The bottom panel of Figure 9 again uses TRAPPIST-1 as the host star with a $CH_4$ flux = $1 \times 10^9$ molecules $cm^{-2}$ $s^{-1}$, now with a $CO_2$ surface volume mixing ratio = 10%. The decreased $CH_4$ predicted with the most conservative prescription leads to smaller $CH_4$ spectral features at 1.7, 2.4, 3.3, and 7.6 $\mu$m, and we still see an $O_3$ feature around 9.6 $\mu$m with the spectrum modeled using the least conservative prescription that is not seen with the spectra modeled using either the most conservative or the recommended prescriptions. Ultimately, the use of either the most conservative or least conservative prescription in models results in spectral differences from those using the recommended prescription; however, the predictions from the most conservative and the recommended prescriptions are most similar. In particular, predicted $O_3$ and CO features are always larger when using the least conservative $CO_2$ dissociation prescriptions.

Figure 10 shows six emission spectra scenarios; on the left, we show emission spectra for (from top to bottom:) Sigma Boötis, the Sun, and TRAPPIST-1 as the host star, for a $CH_4$ flux = $1 \times 10^9$ molecules $cm^{-2}$ $s^{-1}$ and a $CO_2$ surface volume mixing ratio = 3%, and on the right, the three spectra correspond to the example profile plots shown in Figure 2. Overall, we do not see as great a cross-section-dependent difference in these spectra, with the exception of the bottom left panel where we see that the spectrum modeled using the least conservative prescription shows a deeper $CH_4$ feature around 7 μm, as well as an $O_3$ feature that is not present in the spectra that use the most conservative or recommended prescription.

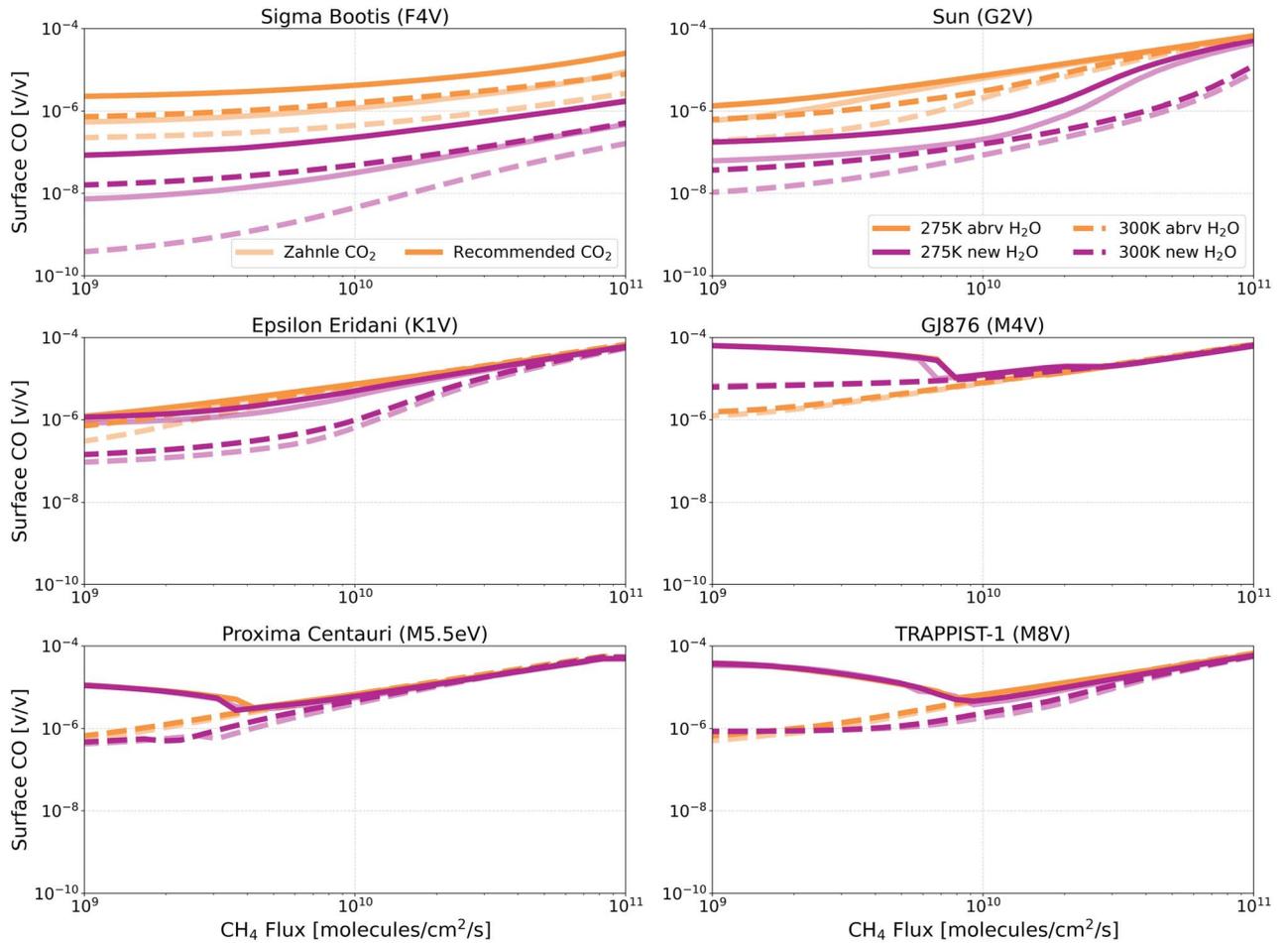

**Figure 13.** Comparison of $H_2O$ cross-section sensitivity tests with old and recommended $CO_2$ cross sections; surface CO vs. $CH_4$ flux for anoxic habitable planets orbiting FGKM-type host stars.

In Figure 11, we show three scenarios for reflection spectra using TRAPPIST-1 as the host star. The top panel shows spectra modeled using a $CH_4$ flux = $1 \times 10^9$ molecules cm$^{-2}$ s$^{-1}$ with the surface $CO_2$ = 3%, the middle panel shows spectra modeled using a $CH_4$ flux = $2 \times 10^{10}$ molecules cm$^{-2}$ s$^{-1}$ with the surface $CO_2$ = 3%, and the bottom panel shows spectra modeled using a $CH_4$ flux = $1 \times 10^9$ molecules cm$^{-2}$ s$^{-1}$ with the surface $CO_2$ = 10%. All three scenarios show differences around 1.7, 2.2, and 2.4 μm due to $CH_4$, as well as a difference due to CO around 2.3 μm.

In general, we find that spectra generated when using the least conservative $CO_2$ cross-section prescription differ the most compared to the spectra generated when using the most conservative and recommended cross-section prescriptions.

### 3.3. Revisiting the $H_2O$ Cross-section Sensitivity Tests

We have revisited the $H_2O$ cross-section sensitivity tests conducted by W. Broussard et al. (2024) with the newly recommended $CO_2$ cross sections, to see how these previous results are impacted. Specifically, we show the impact on the tests conducted using the new $H_2O$ cross sections from S. Ranjan et al. (2020), and the abbreviated versions of these $H_2O$ cross sections, which use a cutoff of 200 nm. Figures 12,

13, and 14 show the impact on the surface $CH_4$, CO, and $O_2$ volume mixing ratios respectively. Broadly, the results from W. Broussard et al. (2024) remain unchanged; terminating the $H_2O$ cross sections at 200 nm results in less $H_2O$ photolysis, thus less OH is produced and trace gases can build up to higher levels, with the cross-section-dependent differences being more pronounced for the FGK-type host stars and negligible for the M-type host stars. By incorporating the recommended $CO_2$ cross sections, we see that the magnitude of the differences has modestly decreased. For example, when using the Zahnle $CO_2$ cross sections to model a planet orbiting the Sun with the 275 K surface temperature regime and a $CH_4$ surface flux of $1 \times 10^{10}$ molecules cm$^{-2}$ s$^{-1}$, the abbreviated $H_2O$ cross sections predict a surface $CH_4$ abundance that is 10 times greater than the surface $CH_4$ abundance predicted when using the new $H_2O$ cross sections. When modeling this same scenario with the recommended $CO_2$ cross sections, the surface $CH_4$ abundance is only 7 times greater when modeled using the abbreviated $H_2O$ cross sections than with the new $H_2O$ cross sections.

Figure 15 shows how key chemical reaction rates have changed with the recommended $CO_2$ cross sections for a habitable anoxic planet orbiting the Sun, for the 275 K surface temperature regime with a $CH_4$ flux of $1.1 \times 10^{10}$ molecules cm$^{-2}$ s$^{-1}$. As expected, with the recommended $CO_2$ cross sections extending further beyond 200 nm, more $CO_2$ photolysis occurs, depleting actinic photons available for H$_2$O photolysis in the troposphere. However, above the troposphere the amount of CO$_2$ photolysis actually decreases with the recommended CO$_2$ cross sections. This is likely because while the recommended CO$_2$ cross sections include temperature-dependent calculations, the Zahnle CO$_2$ cross sections do not. Since temperature decreases above the troposphere, the recommended prescription utilizes the 195 K cross sections, which are smaller than the Zahnle cross sections at wavelengths shorter than around 190 nm.

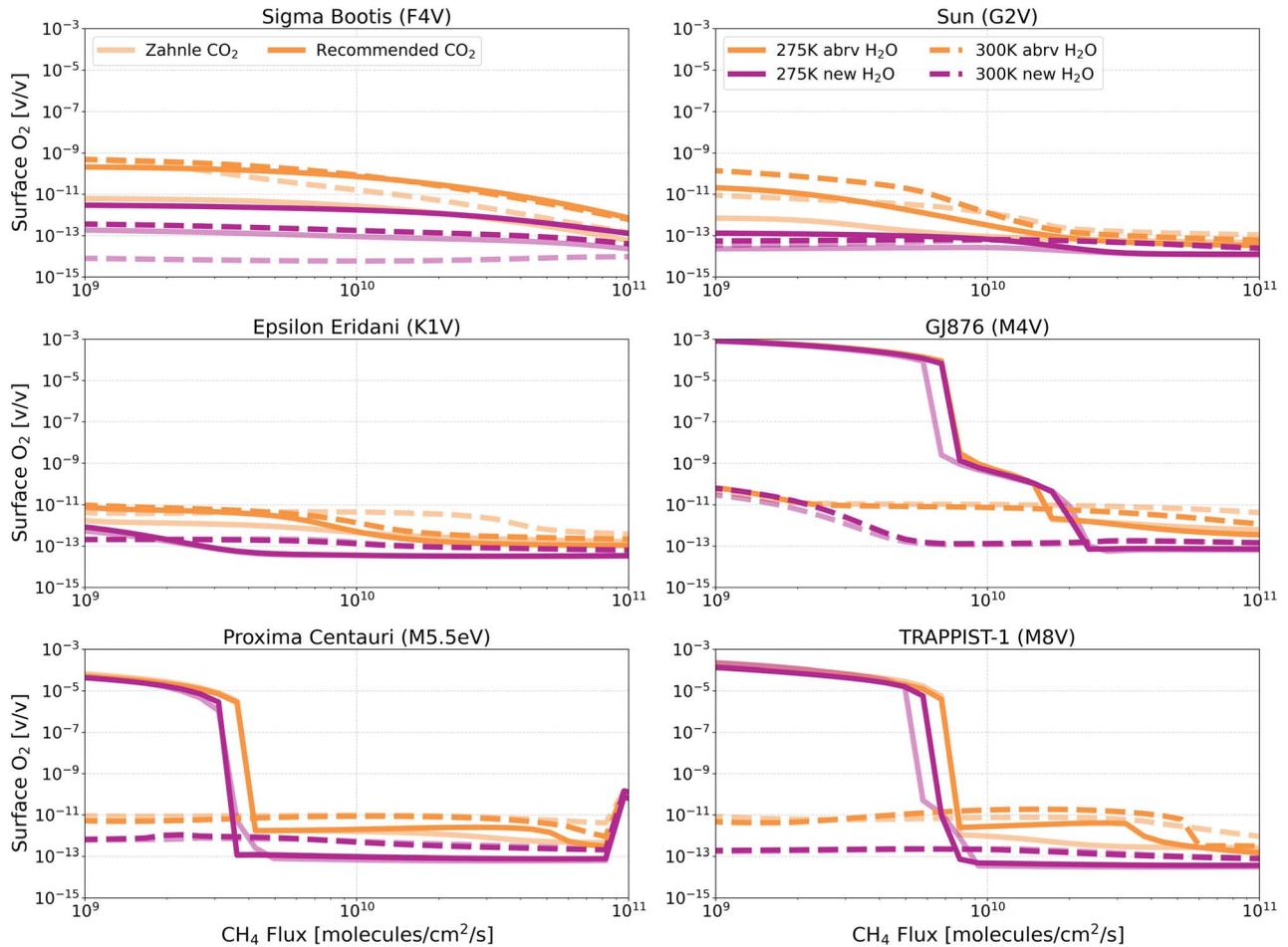

**Figure 14.** Comparison of H$_2$O cross-section sensitivity tests with old and recommended CO$_2$ cross sections; surface O$_2$ vs. CH$_4$ flux for anoxic habitable planets orbiting FGKM-type host stars.

## 4. Discussion

The TRAPPIST-1 planetary system represents a prime opportunity for demonstrating JWST's capability to detect and characterize secondary atmospheres on planets orbiting mid-to-late M-type stars (TRAPPIST-1 JWST Community Initiative et al. 2024). There is currently a strong community focus on studying the TRAPPIST-1 planets, and JWST transmission spectra of these planets will reveal consequential insights into the ability of terrestrial planets orbiting M-type stars to retain their atmospheres (TRAPPIST-1 JWST Community Initiative et al. 2024). As we can see from the transmission spectra modeled in Figure 9, the choice of CO$_2$ cross-section prescription can have a consequential impact on the resulting transmission spectrum. Ruling out our least conservative, highest opacity CO$_2$ prescription could meaningfully affect the interpretation of potential JWST transmission spectra.

A. H. M. J. Triaud et al. (2024) propose the depletion of atmospheric carbon, relative to other planets in the same system, as a potential biosignature. This potential biosignature would be supported by the presence of O$_3$, which could distinguish between a habitable planet and an inhabited one, assuming the O$_3$ is an indirect product of oxygenic photosynthesis. However, we predict that O$_3$ could appear on an abiotic habitable planet assuming our least conservative CO$_2$ prescription, which enhances CO$_2$ photolysis and abiotic O$_2$/O$_3$ production. This could also be problematic for interpreting observations of planets orbiting F-type stars, as the least conservative CO$_2$ cross sections led to a feature of O$_3$ in the emission spectrum for both the scenario of CH$_4$ flux = $1 \times 10^9$ molecules cm$^{-2}$ s$^{-1}$ and that of CH$_4$ flux = $2 \times 10^{10}$ molecules cm$^{-2}$ s$^{-1}$. In other words, this is a possible false positive for the proposed CH$_4$–O$_2$ disequilibrium biosignature (C. Sagan et al. 1993). However, future measurements that conclusively exclude the least conservative CO$_2$ absorption cross sections presented here would preclude these challenging scenarios.

For emission spectra of planets orbiting Sun-like G-type stars with the upcoming HWO, the resulting spectra are more robust against the different predictions based on CO$_2$ cross-section prescription. As shown in Figure 10, with the Sun as a host star and a surface CH$_4$ flux of $10^9$ molecules cm$^{-2}$ s$^{-1}$ the resulting emission spectra show no cross-section-dependent differences. Even when increasing the surface CH$_4$ flux to $2 \times 10^{10}$ molecules cm$^{-2}$ s$^{-1}$, the resulting spectra show a maximum percentage difference of about 45% (corresponding to an absolute difference of ~2 W m$^{-2}$ μm$^{-1}$) around 7.8 μm.

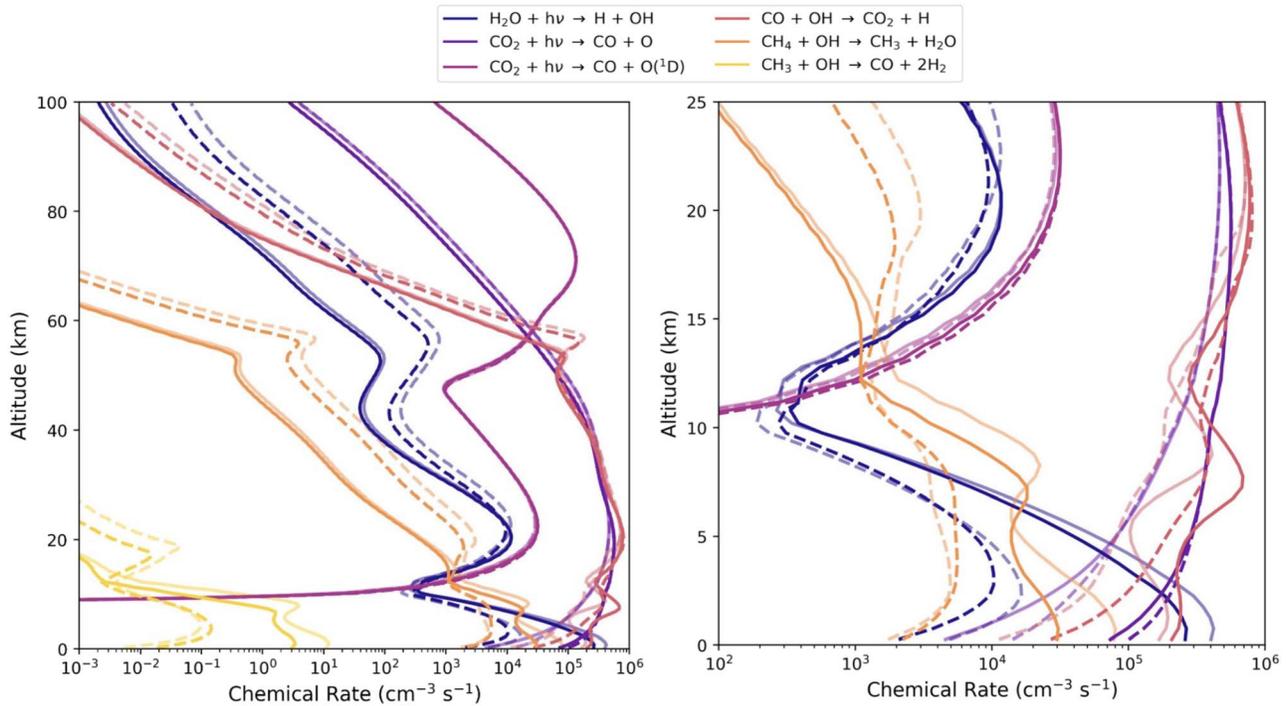

**Figure 15.** Comparison of the reaction rates from the H$_2$O cross-section sensitivity tests with old and recommended CO$_2$ cross sections for a habitable anoxic planet orbiting the Sun, for a CH$_4$ flux of $1.1 \times 10^{10}$ molecules cm$^{-2}$ s$^{-1}$ and a 275 K surface temperature regime. Solid lines are reaction rates modeled using the new H$_2$O cross sections, dashed lines are reaction rates modeled using the abbreviated H$_2$O cross sections; results from W. Broussard et al. (2024) are plotted with 50% opacity, and updated results using the recommended CO$_2$ cross sections are plotted with 100% opacity. On the left, reaction rates up to 100 km are shown; the panel on the right shows the reaction rates up to 25 km.

This research is limited in that it models the impact of extended CO$_2$ cross sections on the trace gas abundances of CH$_4$, CO, and O$_2$ in Archean Earth-like N$_2$–CO$_2$–H$_2$O-dominated atmospheres. This scenario represents just one of the many different possible planetary archetypes we may observe for exoplanets. Our photochemical predictions of other atmospheric scenarios may be more or less impacted by the different CO$_2$ cross-section extrapolations. In particular, the temperature-dependent behavior of the CO$_2$ cross sections may lead to greater cross-section-dependent differences for other planetary scenarios. For example, an oxygen-rich planet with a stratospheric ozone layer will also have an increased stratospheric temperature. Because the least conservative CO$_2$ cross sections are so much larger at higher temperatures, scenarios modeled using this prescription would exhibit even stronger temperature-dependent differences than those modeled with the other prescriptions. Overall, the best way to decrease the uncertainty of our forward and retrieval models for all planetary scenarios would be to obtain additional high-quality CO$_2$ cross-section data at MUV wavelengths.

## 5. Conclusions

We have tested the impact of extended CO$_2$ cross sections (>200 nm) on the atmospheric trace gas abundance of anoxic, temperate, terrestrial exoplanet atmospheres, over a range of CH$_4$ fluxes and CO$_2$ surface mixing ratios for planets orbiting FGKM-type stars. Overall, we can see up to several orders of magnitude in variation of certain trace gas abundances, depending on the cross-section prescription; however, this is with the caveat that the majority of the variation comes in when considering the least conservative prescription and much of this variation is at mixing ratios too small to be observable. If we were able to rule out the least conservative prescription, we would be able to eliminate a large source of the uncertainty in the resulting atmospheric trace gas abundances that could plausibly be spectrally detected. For example, demonstrating that $\sigma_{CO_2} < 1 \times 10^{-24}$ cm$^{-2}$ for $\lambda = 210$–220 nm would strongly falsify the "least conservative" prescription.

The results presented here do not change the conclusions of W. Broussard et al. (2024) in terms of the sensitivity of trace gas abundances in temperate anoxic atmospheres to extended H$_2$O cross sections. Having accurate fundamental modeling inputs is essential for modeling the atmospheres of potentially habitable planets and for being able to correctly interpret observations of terrestrial exoplanets. Repeated measurements of CO$_2$'s cross sections in the MUV at temperatures relevant for habitability—in particular, measurements that are precise enough to rule out the least conservative prescription presented here—will allow us to shed the largest amount of uncertainty and rule out the most incompatible model results.

Overall, this work emphasizes the urgent need for additional laboratory and ab initio studies on fundamental photochemical factors, such as absorption cross sections. Having precise model inputs is essential for interpreting exoplanet spectra, which may eventually reveal atmospheric chemical signatures of life.

## Acknowledgments

This work was performed by the Experimental Constraints for Improving Terrestrial Exoplanet Photochemical Models

(ExCITE-PM) Team funded by NASA Exoplanet Research Program (XRP) grant No. 80NSSC22K0235. W.B. and E.W. S. were further supported by the NASA Interdisciplinary Consortium for Astrobiology Research (ICAR) with funding issued through the Alternative Earths Team (grant No. 80NSSC21K0594) and the Consortium on Habitability and Atmospheres of M-dwarf Planets (CHAMPS) Team (grant No. 80NSSC23K1399). O.V. acknowledges funding from the ANR project "EXACT" (ANR-21-CE49-0008-01) and from the Centre National d'Études Spatiales (CNES). Computations were performed using the computer clusters and data storage resources of the HPCC, which were funded by grants from NSF (MRI-2215705, MRI-1429826) and NIH (1S10OD016290-01A1). We would like to thank the anonymous reviewer for their feedback, which helped us improve this paper.

*Software:* Atmos (G. Arney et al. 2016), SMART (V. S. Meadows & D. Crisp 1996; D. Crisp 1997).

# Appendix A
# Biotic Atmospheric Boundary Conditions

Table 3 lists the biotic atmospheric boundary conditions used for modeling results from the main text of this paper. These are the surface fluxes, surface volume mixing ratios, and dry deposition velocities where relevant.

**Table 3**
Atmospheric Species Boundary Conditions

| Species | Surface Flux (molecules cm$^{-2}$ s$^{-1}$) | Surface Mixing Ratio (v/v) | Dry Deposition Velocity (cm s$^{-1}$) |
|---|---|---|---|
| O($^3$P) | ... | ... | 1 |
| O$_2$ | ... | ... | 0 |
| H$_2$O | ... | ... | 0 |
| H | ... | ... | 1 |
| OH | ... | ... | 1 |
| HO$_2$ | ... | ... | 1 |
| H$_2$O$_2$ | ... | ... | $2 \times 10^{-1}$ |
| H$_2$ | $1 \times 10^{10}$ | ... | $2.4 \times 10^{-4}$ |
| CO | $1 \times 10^{8}$ | ... | $1.2 \times 10^{-4}$ |
| HCO | ... | ... | 1 |
| H$_2$CO | ... | ... | $2 \times 10^{-1}$ |
| CH$_4$ | $10^{9}$–$10^{11}$ | ... | ... |
| CH$_3$ | ... | ... | 1 |
| C$_2$H$_6$ | ... | ... | 0 |
| NO | ... | ... | $3 \times 10^{-4}$ |
| NO$_2$ | ... | ... | $3 \times 10^{-3}$ |
| HNO | ... | ... | 1 |
| O$_3$ | ... | ... | $7 \times 10^{-2}$ |
| HNO$_3$ | ... | ... | $2 \times 10^{-1}$ |
| N | ... | ... | 0 |
| C$_3$H$_2$ | ... | ... | 0 |

**Table 3**
(Continued)

| Species | Surface Flux (molecules cm$^{-2}$ s$^{-1}$) | Surface Mixing Ratio (v/v) | Dry Deposition Velocity (cm s$^{-1}$) |
|---|---|---|---|
| C$_3$H$_3$ | ... | ... | 0 |
| CH$_3$C$_2$H | ... | ... | 0 |
| CH$_2$CCH$_2$ | ... | ... | 0 |
| C$_3$H$_5$ | ... | ... | 0 |
| C$_3$H$_6$ | ... | ... | 0 |
| C$_3$H$_7$ | ... | ... | 0 |
| C$_3$H$_8$ | ... | ... | 0 |
| C$_2$H$_4$OH | ... | ... | 0 |
| C$_2$H$_2$OH | ... | ... | 0 |
| C$_2$H$_5$ | ... | ... | 0 |
| C$_2$H$_4$ | ... | ... | 0 |
| CH | ... | ... | 0 |
| CH$_3$O$_2$ | ... | ... | 0 |
| CH$_3$O | ... | ... | 0 |
| CH$_2$CO | ... | ... | 0 |
| CH$_3$CO | ... | ... | 0 |
| CH$_3$CHO | ... | ... | 0 |
| C$_2$H$_2$ | ... | ... | 0 |
| CH$_2$3 | ... | ... | 0 |
| C$_2$H | ... | ... | 0 |
| C$_2$ | ... | ... | 0 |
| C$_2$H$_3$ | ... | ... | 0 |
| HCS | ... | ... | 0 |
| CS$_2$ | ... | ... | 0 |
| CS | ... | ... | 0 |
| OCS | ... | ... | 0 |
| S | ... | ... | 0 |
| HS | ... | ... | 0 |
| H$_2$S | $3.5 \times 10^{8}$ | ... | $2 \times 10^{-2}$ |
| SO$_3$ | ... | ... | 0 |
| HSO | ... | ... | 1 |
| H$_2$SO$_4$ | ... | ... | 1 |
| SO$_2$ | $3 \times 10^{9}$ | ... | 1 |
| SO | ... | ... | 0 |
| CO$_2$ | ... | $10^{-6}$–$\sim 5 \times 10^{-1}$ | ... |
| SO$_4$AER | ... | ... | 0.01 |
| S$_8$AER | ... | ... | 0.01 |
| HCAER | ... | ... | 0.01 |
| HCAER2 | ... | ... | 0.01 |

# Appendix B
# Alternative Visualization of Model Differences

Figures 16–21 more quantitatively show the differences in atmospheric trace gas abundances for the F-, G-, and K-type host stars, as well as the M5.5eV host star.

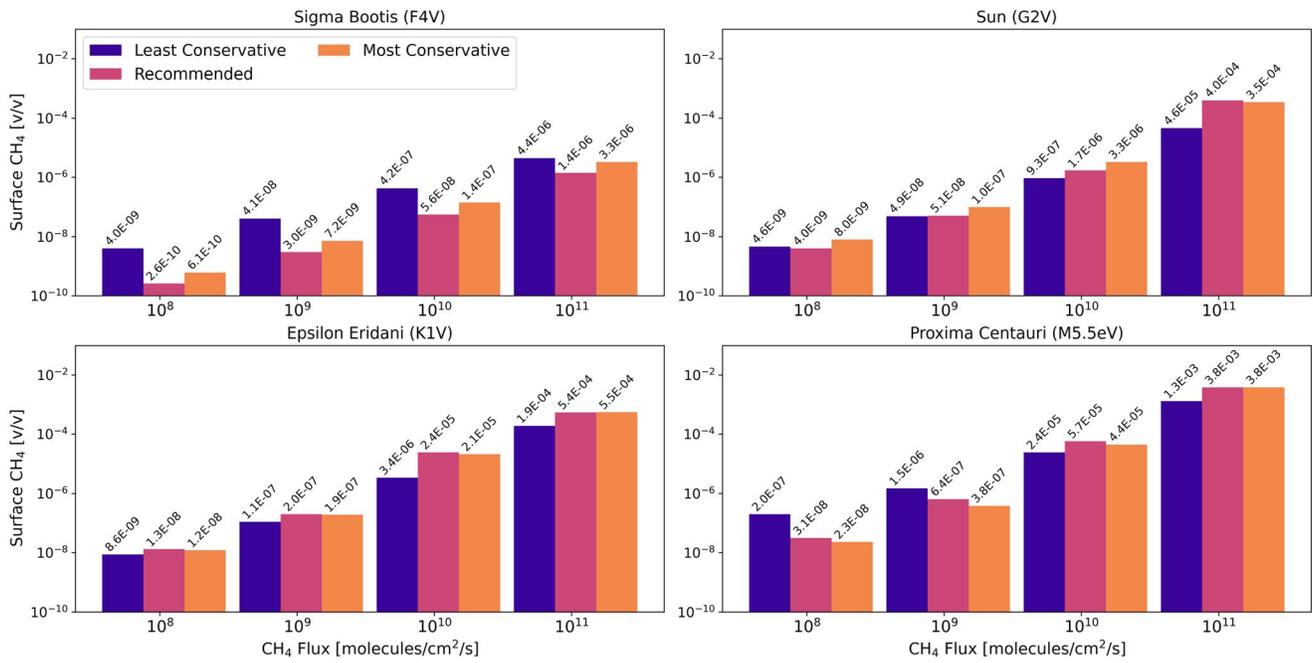

**Figure 16.** Quantitative difference between CH$_4$ surface mixing ratios for CH$_4$ surface fluxes of $10^8$, $10^9$, $10^{10}$, and $10^{11}$ molecules cm$^{-2}$ s$^{-1}$, for anoxic habitable planets modeled using Sigma Boötis, the Sun, Epsilon Eridani, and Proxima Centauri as the host star.

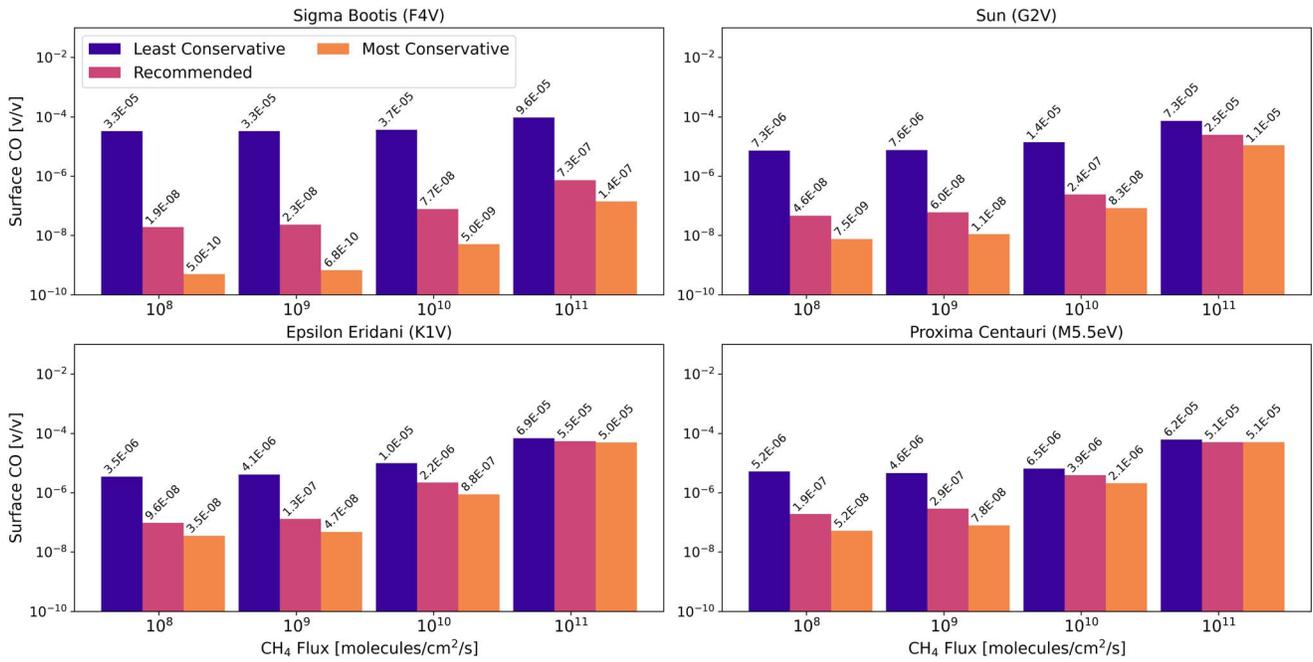

**Figure 17.** Quantitative difference between CO surface mixing ratios for CH$_4$ surface fluxes of $10^8$, $10^9$, $10^{10}$, and $10^{11}$ molecules cm$^{-2}$ s$^{-1}$, for anoxic habitable planets modeled using Sigma Boötis, the Sun, Epsilon Eridani, and Proxima Centauri as the host star.

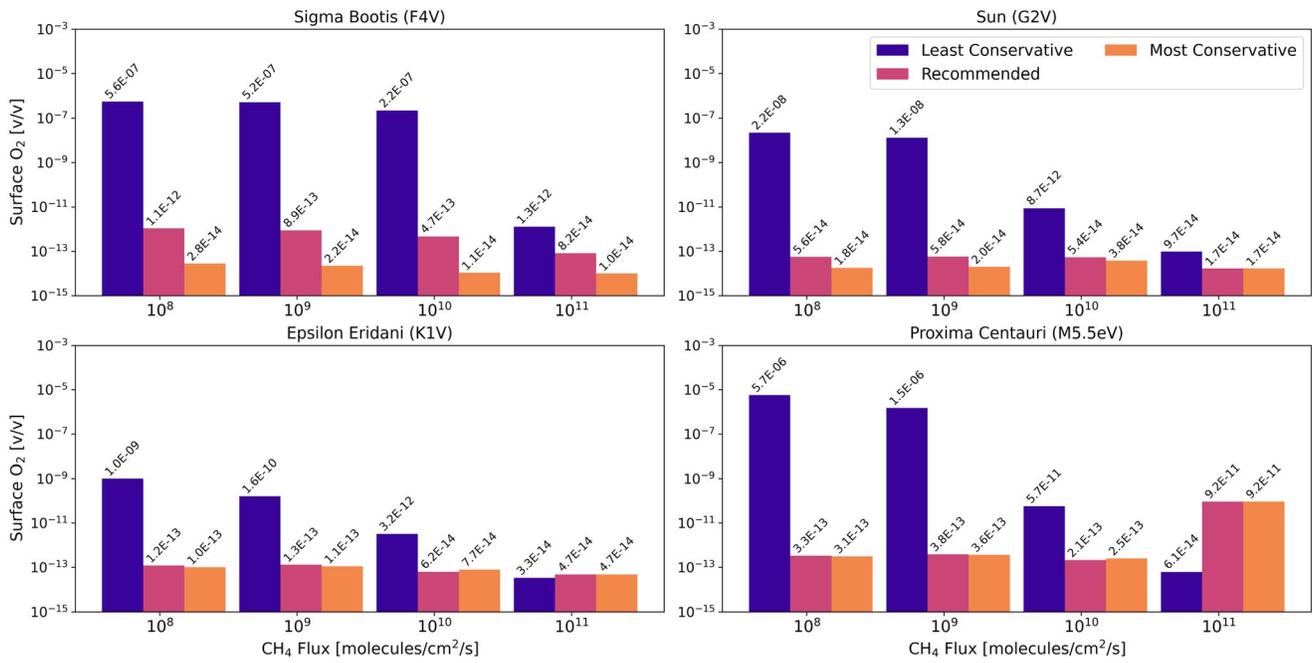

**Figure 18.** Quantitative difference between $O_2$ surface mixing ratios for $CH_4$ surface fluxes of $10^8$, $10^9$, $10^{10}$, and $10^{11}$ molecules cm$^{-2}$ s$^{-1}$, for anoxic habitable planets modeled using Sigma Boötis, the Sun, Epsilon Eridani, and Proxima Centauri as the host star.

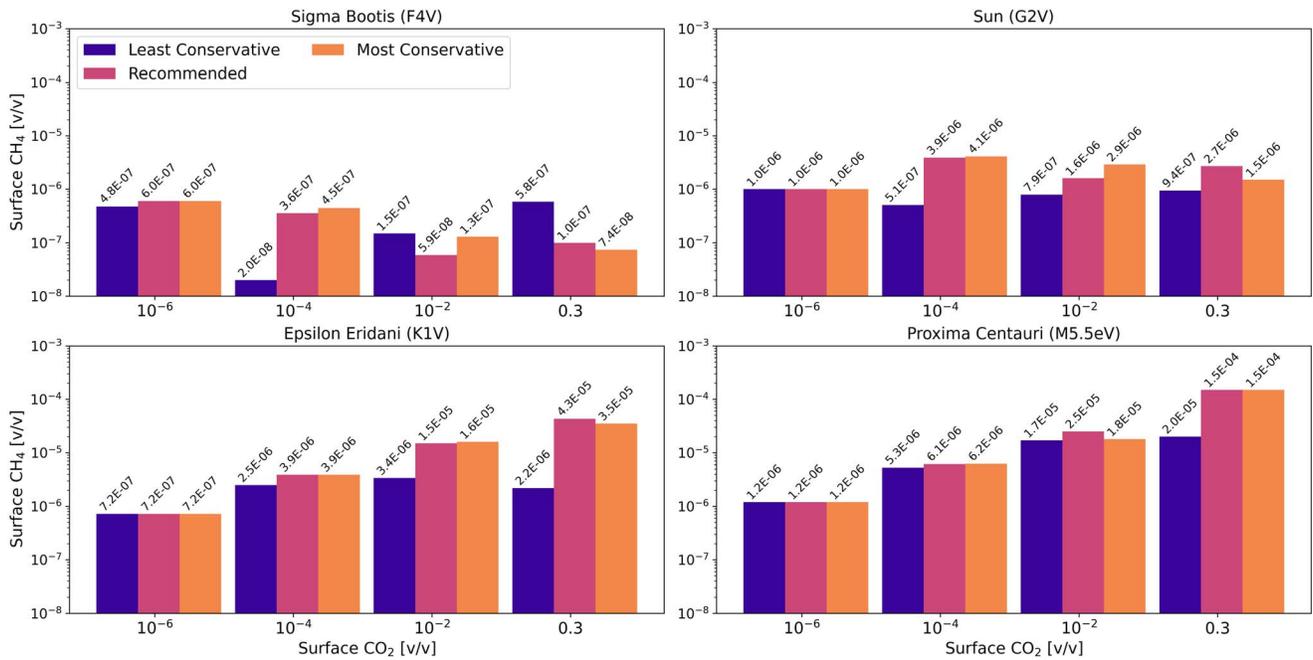

**Figure 19.** Quantitative difference between $CH_4$ surface mixing ratios for $CO_2$ surface mixing ratios of $10^{-6}$, $10^{-4}$, $10^{-2}$, and 0.3 (v/v), for anoxic habitable planets modeled using Sigma Boötis, the Sun, Epsilon Eridani, and Proxima Centauri as the host star.

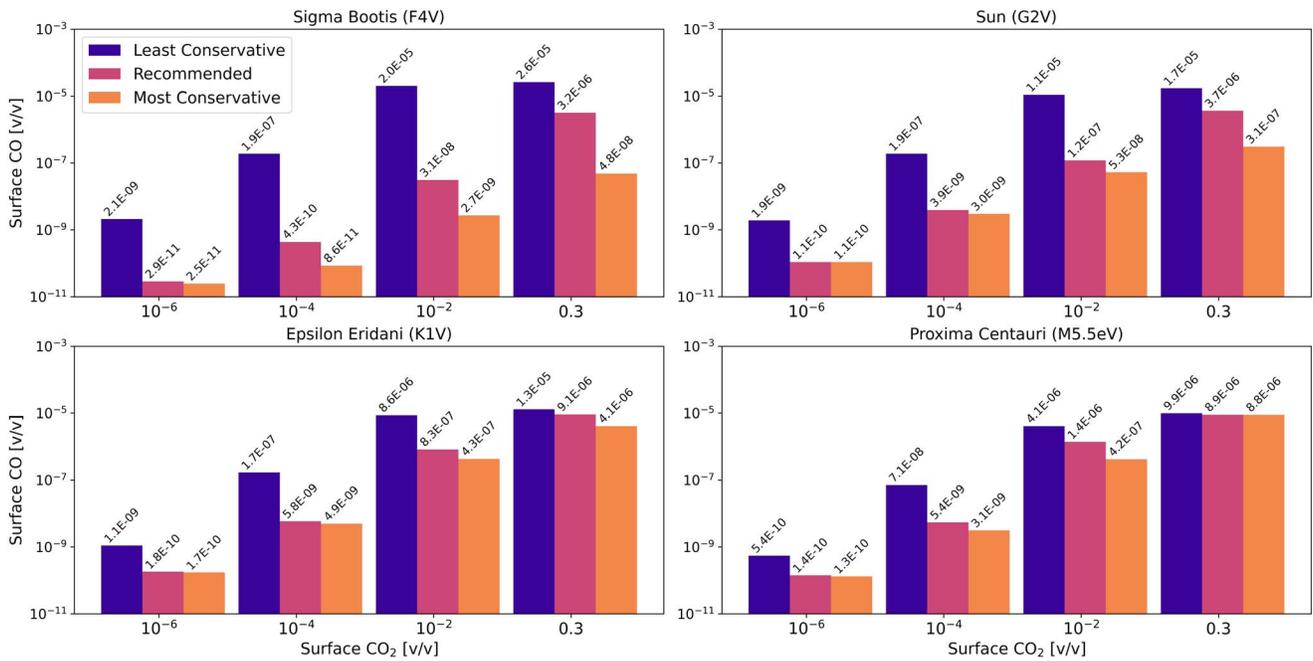

**Figure 20.** Quantitative difference between CO surface mixing ratios for $CO_2$ surface mixing ratios of $10^{-6}$, $10^{-4}$, $10^{-2}$, and 0.3 (v/v), for anoxic habitable planets modeled using Sigma Boötis, the Sun, Epsilon Eridani, and Proxima Centauri as the host star.

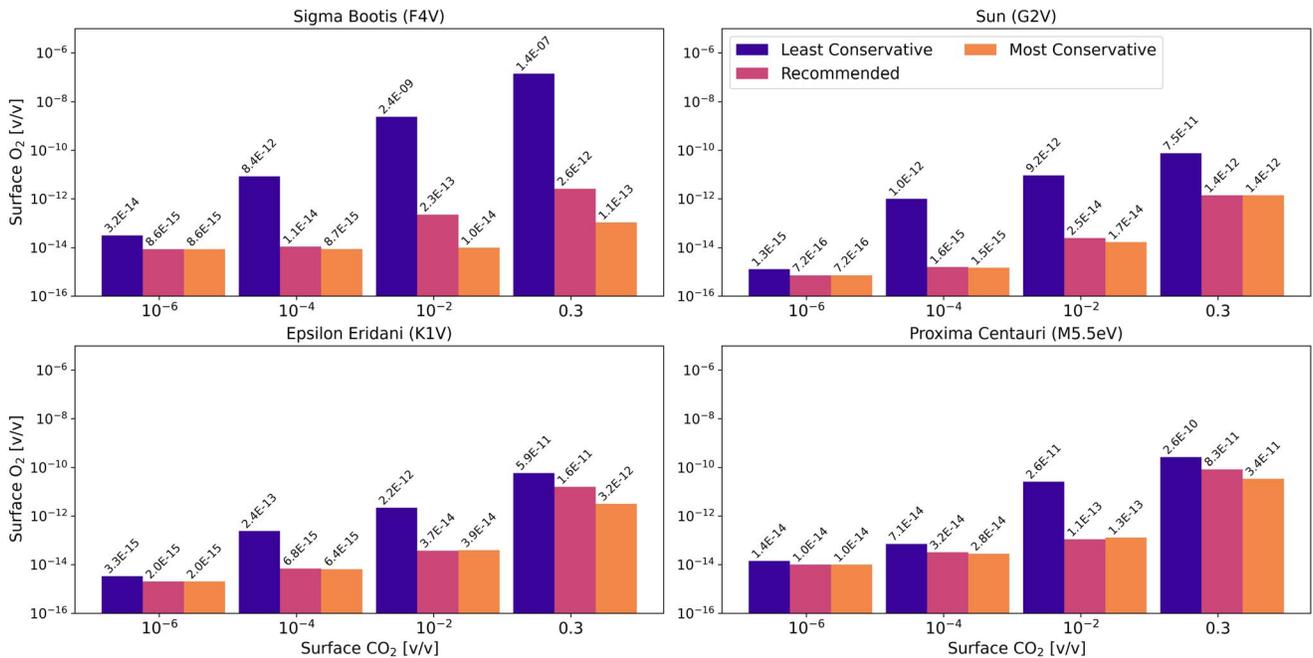

**Figure 21.** Quantitative difference between $O_2$ surface mixing ratios for $CO_2$ surface mixing ratios of $10^{-6}$, $10^{-4}$, $10^{-2}$, and 0.3 (v/v), for anoxic habitable planets modeled using Sigma Boötis, the Sun, Epsilon Eridani, and Proxima Centauri as the host star.

## ORCID iDs

Wynter Broussard ● https://orcid.org/0009-0006-2304-3419
Edward W. Schwieterman ● https://orcid.org/0000-0002-2949-2163
Clara Sousa-Silva ● https://orcid.org/0000-0002-7853-6871
Grace Sanger-Johnson ● https://orcid.org/0000-0002-4463-2902
Sukrit Ranjan ● https://orcid.org/0000-0002-5147-9053
Olivia Venot ● https://orcid.org/0000-0003-2854-765X

## References

Arney, G., Domagal-Goldman, S. D., Meadows, V. S., et al. 2016, AsBio, 16, 873
Arney, G., Domagal-Goldman, S. D., & Meadows, V. S. 2018, AsBio, 18, 311
Barth, P., Stüeken, E. E., Helling, C., Schwieterman, E. W., & Telling, J. 2024, A&A, 686, A58
Broussard, W., Schwieterman, E. W., Ranjan, S., et al. 2024, ApJ, 967, 114
Crisp, D. 1997, GeoRL, 24, 571
Currie, M. H., Meadows, V. S., & Rasmussen, K. C. 2023, PSJ, 4, 83
Del Genio, A. D., Kiang, N. Y., Way, M. J., et al. 2019, ApJ, 884, 75

Felton, R. C., Bastelberger, S. T., Mandt, K. E., et al. 2022, JGRE, 127, e2021JE006853
France, K., Loyd, R. O. P., Youngblood, A., et al. 2016, ApJ, 820, 89
Gaillard, F., & Scaillet, B. 2014, E&PSL, 403, 307
Gao, P., Hu, R., Robinson, T. D., Li, C., & Yung, Y. L. 2015, ApJ, 806, 249
Greene, T. P., Line, M. R., Montero, C., et al. 2016, ApJ, 817, 17
Harman, C., Schwieterman, E., Schottelkotte, J. C., & Kasting, J. 2015, ApJ, 812, 137
Hu, R., Damiano, M., Scheucher, M., et al. 2021, ApJL, 921, L8
Ityaksov, D., Linnartz, H., & Ubachs, W. 2008, CPL, 462, 31
Kaltenegger, L. 2017, ARA&A, 55, 433
Kasting, J. F., & Ackerman, T. P. 1986, Sci, 234, 1383
Kempton, E. M. R., & Knutson, H. A. 2024, RvMG, 90, 411
Kopparapu, R. K., Ramirez, R., Kasting, J. F., et al. 2013, ApJ, 765, 131
Krissansen-Totton, J., Garland, R., Irwin, P., & Catling, D. C. 2018, AJ, 156, 114
Lincowski, A. P., Meadows, V. S., Crisp, D., et al. 2018, ApJ, 867, 76
Loyd, R. O. P., France, K., Youngblood, A., et al. 2016, ApJ, 824, 102
Loyd, R. O. P., Shkolnik, E. L., Schneider, A. C., et al. 2018, ApJ, 867, 70
Madhusudhan, N., Moses, J. I., Rigby, F., & Barrier, E. 2023, FaDi, 245, 80
Mamajek, E., & Stapelfeldt, K. 2024, AAS Meeting, 56, 628.17
May, E. M., MacDonald, R. J., Bennett, K. A., et al. 2023, ApJL, 959, L9
Meadows, V. S., & Crisp, D. 1996, JGR, 101, 4595
Meadows, V. S., Lincowski, A. P., & Lustig-Yaeger, J. 2023, PSJ, 4, 192
Morley, C. V., Kreidberg, L., Rustamkulov, Z., Robinson, T., & Fortney, J. J. 2017, ApJ, 850, 121
National Academies of Sciences, Engineering, and Medicine 2021, Pathways to Discovery in Astronomy and Astrophysics for the 2020s (Washington, DC: National Academies Press)
Parkinson, W., Rufus, J., & Yoshino, K. 2003, CP, 290, 251
Peacock, S., Barman, T., Shkolnik, E. L., Hauschildt, P. H., & Baron, E. 2019a, ApJ, 871, 235
Peacock, S., Barman, T., Shkolnik, E. L., et al. 2019b, ApJ, 886, 77
Peacock, S., Barman, T., Shkolnik, E. L., et al. 2020, ApJ, 895, 5
Ranjan, S., Schwieterman, E. W., Harman, C., et al. 2020, ApJ, 896, 148
Sagan, C., Thompson, W. R., Carlson, R., Gurnett, D., & Hord, C. 1993, Natur, 365, 715
Sander, S. P., Abbatt, J., Barker, J. R., et al. 2011, Chemical Kinetics and Photochemical Data for Use in Atmospheric Studies Evaluation Number 17, NASA JPL, https://jpldataeval.jpl.nasa.gov/
Schmidt, J. A., Johnson, M. S., & Schinke, R. 2013, PNAS, 110, 17691
Schwieterman, E. W., Kiang, N. Y., Parenteau, M. N., et al. 2018, AsBio, 18, 663
Schwieterman, E. W., Meadows, V. S., Domagal-Goldman, S. D., et al. 2016, ApJL, 819, L13
Schwieterman, E. W., Olson, S. L., Pidhorodetska, D., et al. 2022, ApJ, 937, 109
Segura, A., Krelove, K., Kasting, J. F., et al. 2003, AsBio, 3, 689
Segura, A. A., Kasting, J. F., Meadows, V., et al. 2005, AsBio, 5, 706
Shkolnik, E. L., & Barman, T. S. 2014, AJ, 148, 64
Thompson, M. A., Krissansen-Totton, J., Wogan, N., Telus, M., & Fortney, J. J. 2022, PNAS, 119, e2117933119
Thuillier, G., Floyd, L., Woods, T. N., et al. 2004, Solar Irradiance Reference Spectra (Washington D. C.: AGU), 171
TRAPPIST-1 JWST Community Initiative, de Wit, J., Doyon, R., et al. 2024, NatAs, 8, 810
Triaud, A. H. M. J., de Wit, J., Klein, F., et al. 2024, NatAs, 8, 17
Venot, O., Bénilan, Y., Fray, N., et al. 2018, A&A, 609, A34
Wen, J.-S., Pinto, J. P., & Yung, Y. L. 1989, JGR, 94, 957
Wogan, N. F., Batalha, N. E., Zahnle, K. J., et al. 2024, ApJL, 963, L7
Youngblood, A., France, K., Loyd, R. O. P., et al. 2016, ApJ, 824, 101